\documentclass[a4paper,12pt]{article}
\pdfoutput=1 
\usepackage{jheppub} 
\usepackage[T1]{fontenc} 
\usepackage{color}

\newcommand{\be}{\begin{equation}}
\newcommand{\ee}{\end{equation}}
\newcommand{\beal}{\begin{aligned}}
\newcommand{\eeal}{\end{aligned}}
\newcommand\bea {\begin{eqnarray}}
\newcommand\eea {\end{eqnarray}}
\newcommand{\bec}{\begin{cases}}
\newcommand{\eec}{\end{cases}}
\newcommand{\bei}{\begin{itemize}}
\newcommand{\eei}{\end{itemize}}
\newcommand{\bee}{\begin{enumerate}}
\newcommand{\eee}{\end{enumerate}}

\newcommand{\beq}{\begin{equation}}
\newcommand{\eeq}{\end{equation}}
\newcommand\bqa {\begin{eqnarray}}
\newcommand\eqa {\end{eqnarray}}

\newcommand\pr {\partial}
\newcommand{\fr}{\frac}

\newcommand\nn {\nonumber}

\newcommand{\bear}{\begin{array}}
\newcommand{\enar}{\end{array}}

\newcommand{\W}{\mathcal{W}}

\newcommand{\tit}{\textit}
\newcommand{\tbf}{\textbf}

\def\I{{\rm i}}

\def\t{\theta}

\def\a{\alpha}
\def\b{\beta}

\def\l{\lambda}
\def\wt{\widetilde}
\def\t{\tilde}

\title{Lifshitz flows in IIB and dual field theories}

\author[a]{Philipp Burda\footnote{On leave of absence from ITEP, Moscow.}}
\author[a,b]{Ruth Gregory}
\author[a]{Simon F.\ Ross}

\affiliation[a]{Centre for Particle Theory, Durham University,
South Road, Durham, DH1 3LE, UK}
\affiliation[b]{Perimeter Institute, 31 Caroline Street North, Waterloo, 
ON, N2L 2Y5, Canada}

\emailAdd{philipp.burda@durham.ac.uk}
\emailAdd{r.a.w.gregory@durham.ac.uk}
\emailAdd{s.f.ross@durham.ac.uk}

\abstract{
We construct solutions describing flows between AdS and 
Lifshitz spacetimes in IIB supergravity. We find that flows 
from AdS$_5$ can approach either AdS$_3$ or Lifshitz$_3$ 
in the IR depending on the values of the deformation from 
AdS$_5$. Surprisingly, the choice between AdS and Lifshitz 
IR depends only on the value of the deformation, not on its 
character; the breaking of the Lorentz symmetry in the flows 
with Lifshitz IR is spontaneous. We find that the values of the 
deformation which lead to flows to Lifshitz make the UV field 
theory dual to the AdS$_5$ geometry unstable, so that these 
flows do not offer an approach to defining the field theory 
dual to the Lifshitz spacetime. 
}

\preprint{DCPT-14/31}

\begin{document}

\maketitle 

\section{Introduction}

The extension of holography \cite{Maldacena:1997zz}
to field theories with dynamical exponent $z >1$ is interesting 
both for the potential application of these theories in condensed 
matter physics and for its potential to enlarge our understanding 
of holographic dualities (for reviews see 
e.g.\ \cite{Sachdev:2010ch,McGreevy:2009xe,Hartnoll:2009sz}).
Such theories have a symmetry under the scaling $t \to \lambda^z t$, 
$\vec x \to \lambda \vec x$, and it was realized in 
\cite{Kachru:2008yh} that a holographic dual could be 
constructed by considering spacetimes with a metric 
\begin{equation}
ds^2 = r^{2z} dt^2 - r^2 d \vec{x}^2 - \frac{dr^2}{r^2}, 
\end{equation}
which have an isometry under $t \to \lambda^z t$, 
$\vec x \to \lambda \vec x$, $r \to \lambda^{-1} r$. 
In \cite{Kachru:2008yh,Taylor:2008tg} simple ``bottom-up'' 
models admitting such solutions were proposed. They have 
since been realized as solutions in ``top-down'' models 
obtained from string theory: the case $z=2$ proves to be 
the simplest to realize 
\cite{Balasubramanian:2010uk,Donos:2010tu,Cassani:2011sv,Dey:2013oba}, 
but a construction allowing for general values of $z$ was given 
in \cite{Gregory:2010gx}. Some other particular values of $z$ 
were also realized in \cite{Singh:2010zs,Singh:2012un,Donos:2010ax}.

An interesting goal in such top-down constructions is to get a better 
understanding of the non-relativistic field theories dual to such 
Lifshitz solutions. It is particularly interesting to understand 
these holographic theories, as no examples of interacting 
theories with Lifshitz symmetries are known. In 
\cite{Braviner:2011kz}, holographic RG flows relating 
the Lifshitz and AdS solutions in the context of the massive 
IIA setup in \cite{Gregory:2010gx} were constructed, and 
it was noted that the RG flows offered a potential approach 
to understanding the field theory dual to Lifshitz, as one 
could consider the flow from an AdS solution with a known 
dual to Lifshitz. Related work on such flows and their applications 
includes \cite{Singh:2010cj,Bertoldi:2010ca,Gubser:2009cg,Goldstein:2010aw,
Hartnoll:2010gu,Liu:2012wf,Singh:2013iba,Kachru:2013voa,Dey:2014yra}. 
A dynamical interpolation was studied in \cite{Uzawa:2013koa}. 
A different approach to relating AdS to Lifshitz is 
\cite{Korovin:2013nha,Korovin:2013bua}. 

In this paper, we extend the work of \cite{Braviner:2011kz} by 
considering flows involving the type IIB Lifshitz solutions in 
\cite{Gregory:2010gx}. We start with the five-dimensional 
gauged supergravity obtained by compactifying IIB on an 
$S^5$, and consider further compactifying two spatial 
directions on a compact hyperbolic space, with certain 
gauge fluxes turned on on this space. There are asymptotically 
AdS$_5$ solutions, where the proper size of compact 
hyperbolic space grows near the boundary, and AdS$_3$ 
and 3-dimensional Lifshitz(denoted Li$_3$) solutions 
where it has constant size. 
As in \cite{Braviner:2011kz}, we consider flows relating 
all these solutions. We focus particularly on the flows 
from AdS$_5$, and analyze these in detail, identifying 
the deformation of AdS$_5$ which source the flow and 
discussing its dual field theory description. 

Working in the IIB context has two advantages: the field 
theory dual to the asymptotically AdS$_5$ solution is the 
familiar $\mathcal N =4$ SYM, and the deformation we are 
interested in includes as a special case a supersymmetric 
twist which has been previously studied in \cite{Maldacena:2000mw}. 
In the supersymmetric flow, \cite{Maldacena:2000mw} showed 
that the twist involves not only turning on a flux $Q$ but also adding a 
source $\lambda$ for a scalar operator transforming in the 
$\mathbf{20}$ of the $SU(4)$ R-symmetry. We will see that 
the flows to non-supersymmetric AdS$_3$ and Lifshitz 
geometries involve changing the values of $Q$ and 
$\lambda$ in a coordinated way: the flow reaches an IR 
fixed point on one-dimensional subspaces in the space of 
$\{Q, \lambda\}$ deformations. 

Surprisingly, we do not need to turn on a source which 
breaks Lorentz symmetry explicitly in the UV to realize 
flows to Lifshitz: this Lorentz symmetry breaking will 
emerge spontaneously for appropriate values of $\{Q,\lambda\}$. 

In \cite{Maldacena:2000mw}, the deformation by $\{Q,\lambda\}$ 
was related to a change in the scalar Lagrangian 
in the $\mathcal N  =4$ SYM theory, and it was shown to 
lead to flat directions for certain scalars in the supersymmetric 
case. We analyze this field theory Lagrangian deformation 
for our non-supersymmetric cases and find that there is a 
finite range of non-supersymmetric flows to AdS$_3$ 
where the flat directions get lifted and the field theory 
scalars in the deformed field theory will be stable in the UV. 
Disappointingly, for the flows to Li$_3$, the field theory 
deformation always leads to some runaway directions in 
the scalar space. These runaways correspond to brane 
nucleation instabilities in the bulk geometry (discussed 
for example in \cite{Seiberg:1999xz,Witten:1999xp}), 
as we show explicitly by a probe brane calculation. 
Thus, for the flows to Lifshitz, the UV field theory is unstable, 
and this flow does not offer us a way to define the IR theory 
dual to the Lifshitz geometry. As in \cite{Braviner:2011kz}, 
we also find that for some values of $z$ the Lifshitz geometries 
have linearized modes which appear to violate the generalization 
of the Breitenlohner-Freedman bound \cite{Breitenlohner:1982jf}. 
These two types of instabilities do not appear to be related. 

In section \ref{model}, we review the Romans 5D gauged
SUGRA model \cite{Romans:1985ps} and review the Lifshitz 
solutions in this model \cite{Gregory:2010gx}, as well
as discussing the families of AdS$_3$ solutions.
We then discuss the flows in section \ref{flows}, first 
performing a linearized analysis about each 
of the solutions to determine the qualitative character of the 
flows and then numerically constructing the various flows. 
In section \ref{stab}, we analyze the deformation away from 
AdS$_5$ in the UV and discuss the dual field theory. 

\section{Lifshitz and AdS solutions in five-dimensional gauged supergravity}
\label{model}

We consider a consistent truncation of the $\mathcal{N}=4$ 
five-dimensional gauged supergravity theory obtained by 
reduction of the ten-dimensional type IIB supergravity 
on $S^5$, where we keep an $SU(2) \times U(1)$ subgroup 
of the $SU(4)$ gauge group, and a single scalar $\phi$ 
\cite{Romans:1985ps}. This theory is a consistent 
truncation of the full higher dimensional theory, in the sense
that any solutions in the 5D theory can be uplifted to 
Type IIB supergravity solutions in ten dimensions 
(see \cite{Lu:1999bw} for explicit detail).

The field content of the theory consists of the metric 
$g_{\mu\nu}$, 5D dilaton field $\phi$, $SU(2)$ gauge field 
$A_{\mu}^{(i)}$, $U(1)$ gauge field $\mathcal{A}_{\mu}$ 
and two antisymmetric tensor fields $B_{\mu\nu}^{\a}$. 
The bosonic part of the Lagrangian is
\be
\beal
\mathcal{L} = &-\fr{R}{4} 
+ \fr12 \partial_{\mu}\phi \partial^{\mu}\phi 
- \fr14 \xi^{-4} \mathcal{F}_{\mu\nu}\mathcal{F}^{\mu\nu} 
- \fr14 \xi^2 \left( F^{(i)}_{\mu\nu}F^{\mu\nu (i)}
+B^{\mu\nu\a}B^{\a}_{\mu\nu} \right) \\
& + \fr14 \epsilon^{\mu\nu\rho\sigma\lambda} 
\left( \fr{1}{g_1} \epsilon_{\a\b} B^{\a}_{\mu\nu}
D_{\rho}B^{\b}_{\sigma\lambda} 
- F^{(i)}_{\mu\nu}F^{(i)}_{\rho\sigma}\mathcal{A}_{\lambda}\right) 
+ P(\phi),
\eeal
\label{TypeIIB-Lagrangian}
\ee
where $\xi = e^{\sqrt{\fr23}\phi}$, the scalar field potential is
\be
P(\phi) = \fr{g_2}{8} \left(g_2 \xi^{-2} + 2\sqrt{2} g_1 \xi \right),
\ee
and field strengths are
\be
\beal
\mathcal{F}_{\mu\nu} & =  \pr_{\mu}\mathcal{A}_{\nu} 
- \pr_{\nu}\mathcal{A}_{\mu}, \\
F^{(i)}_{\mu\nu} & =  \pr_{\mu}A^{(i)}_{\nu} 
- \pr_{\nu}A^{(i)}_{\mu} + g_2 \epsilon^{ijk} A^{(j)}_{\mu} A^{(k)}_{\nu}. 
\eeal
\ee
The $U(1)$ gauge coupling $g_1$ and $SU(2)$ gauge 
coupling $g_2$ are two independent parameters of the theory. 
It was shown in \cite{Romans:1985ps} that these parameters 
can be eliminated by field redefinitions so that there are only 
three physically different theories, the $\mathcal{N}=4^{+}$ 
theory, when $g_1g_2 > 0$,  the $\mathcal{N}=4^{0}$ theory, 
when $g_2 = 0$, and the $\mathcal{N}=4^{-}$ theory, when 
$g_1g_2 < 0$. We will consider here only the $\mathcal{N}=4^{+}$ 
theory, i.e.\ we assume $g_1g_2 > 0$. We also set 
$B_{\mu\nu}^{\a} = 0$ identically for all solutions and flows 
considered here. 

The equations of motion for the rest of the fields are then
\be
\beal
R_{\mu\nu} & =  2\pr_{\mu}\phi\pr_{\nu}\phi 
+ \fr43 g_{\mu\nu} P(\phi) 
- \xi^{-4} \left( 2\mathcal{F}_{\mu\rho}\mathcal{F}_{\nu}^{\rho} 
- \fr13 g_{\mu\nu} \mathcal{F}_{\rho\sigma}\mathcal{F}^{\rho\sigma} 
\right)  \\
&\quad - \xi^2 \left( 2 F^{(i)}_{\mu\rho}F_{\nu}^{\rho (i)} 
- \fr13 g_{\mu\nu} F^{(i)}_{\rho\sigma}F^{\rho\sigma (i)}\right), \\
\square \phi & =  \fr{\pr P}{\pr \phi} 
+ \sqrt{\fr23} \xi^{-4} \mathcal{F}_{\mu\nu}\mathcal{F}^{\mu\nu} 
- \sqrt{\fr16} \xi^2 F^{(i)}_{\rho\sigma}F^{(i) \rho\sigma}, \\
D_{\nu} \left( \xi^{-4} \mathcal{F}^{\nu\mu} \right) 
& =  \fr14 \epsilon^{\mu\nu\rho\sigma\tau} 
F^{(i)}_{\nu\rho} F^{(i)}_{\sigma\tau}, \\
D_{\nu} \left( \xi^2  F^{\nu\mu (i)} \right) 
& =  \fr12 \epsilon^{\mu\nu\rho\sigma\tau} 
F^{(i)}_{\nu\rho} \mathcal{F}_{\sigma\tau}. 
\eeal
\label{TypeIIB-eoms}
\ee

\subsection{Ansatz for solutions and flows}

To construct flows, we only need to consider radial dependence 
of the bulk fields; we assume the holographic RG flow geometries 
we consider will preserve the translational invariance in the $t$ 
and $x$ directions, and will have the topological flux through the
compact hyperbolic space. The most general ansatz we will need to 
consider is thus 
\be
ds^2 = e^{2F(r)} dt^2 - r^2 dx^2 - e^{2d(r)} \fr{dr^2}{r^2} 
- e^{2h(r)} \fr{dy_{1}^2 + dy_2^2}{y_2^2}, 
\label{metric}
\ee
the 5D dilaton $\phi$ is also only a function of $r$, and we assume 
the gauge fields have at most nonzero $r-t$ or $r-x$ components.
It is convenient to parametrize the fields in such a way as to eliminate
geometric factors:
\be
\beal
F^{(3)}_{rt} & = \frac{{\tilde{A}}(r)}{\xi r} e^{F+D}\;,&\quad
F^{(3)}_{rx} & =  \frac{B(r)}{\xi}e^{D}\;,&\quad
F^{(3)}_{y_1y_2} & =  \fr{Q}{g_2 y_2^2}\;,\\
\mathcal{F}_{rt} &= \frac{A(r) \xi^2}{r} e^{F+ D}\;, 
&\quad
\mathcal{F}_{rx} &= {\tilde{B}(r)} \xi^2 e^{D}\;,& \\
\eeal
\label{gauge}
\ee
where we have also introduced shifted and rescaled variables in 
order to eliminate $g_1$ and $g_2$ from all expressions:
\be
\beal
D(r) & = d(r) + \fr13 \ln\left(g_1g_2^2\right), \\
H(r) & = h(r) + \fr13 \ln\left(g_1g_2^2\right), \\
\varphi(r) & = \xi^3(r) g_1 g_2^{-1}, 
\eeal
\ee
Substituting all this into the equations (\ref{TypeIIB-eoms}) and 
introducing the new variable $\rho=\ln{r}$ we get
\be
\beal
\frac{R^t_t}{g_1^{\fr23} g_2^{\fr43}} & = e^{-2D} \left[
F' - F'D' + F^{\prime 2} + F'' + 2 H'F' \right] \\
& = \fr16 \left(\varphi^{-\fr23} + 2\sqrt{2}\varphi^{\fr13}\right) 
+ \fr43 \left(A^2 + \t{A}^2 \right) + \fr23 \left(\t{B}^2 + B^2 \right) 
+ \fr23 \varphi^{\fr23} Q^2 e^{-4H} \\
\frac{R^x_x}{g_1^{\fr23} g_2^{\fr43}} & = e^{-2D} \left[
F' - D' + 1 + 2 H' \right]\\
& = \fr16 \left(\varphi^{-\fr23} + 2\sqrt{2}\varphi^{\fr13}\right)  
- \fr23 \left(A^2 + \t{A}^2 \right) - \fr43 \left(\t{B}^2 + B^2 \right) 
+ \fr23 \varphi^{\fr23} Q^2 e^{-4H} \\
\frac{R^r_r}{g_1^{\fr23} g_2^{\fr43}}  &=  e^{-2D} \left[
F'' + F^{\prime 2} - F' D' - D' +1 - 2 H' D' + 2 H^{\prime 2}
+ 2 H'' \right ] \\
& = \frac{-\varphi^{\prime 2}}{3\varphi^2e^{2D}} 
+ \fr16 \left(\varphi^{-\fr23} + 2^{\fr32}\varphi^{\fr13}\right) 
+ \fr43 \left(A^2 + \t{A}^2 - \t{B}^2 - B^2 \right) 
+  \fr23 \varphi^{\fr23} Q^2 e^{-4H} \\
\frac{R^{y_1}_{y_1}}{g_1^{\fr23} g_2^{\fr43}}
& = e^{-2H} + e^{-2D}\left[
H'' + 2 H^{\prime 2} + H'F' + H' - H'D' \right]\\
& = \fr16 \left(\varphi^{-\fr23} + 2\sqrt{2}\varphi^{\fr13}\right) 
- \fr23 \left(A^2 + \t{A}^2 \right) + \fr23 \left(\t{B}^2 + B^2 \right) 
- \fr43 \varphi^{\fr23} Q^2 e^{-4H} 
\eeal
\label{einstein-eqns}
\ee
for the Einstein equations, where a prime now denotes $\pr_\rho$,
and
\be
\beal
\square \ln\varphi & = - e^{-2D} \pr_{\rho}^2\ln\varphi 
- e^{-2D}\pr_{\rho}\ln\varphi \left(1 + F' - D' + 2H' \right)  \\
& = \fr12 \left(-\varphi^{-\fr23} + \sqrt{2}\varphi^{\fr13}\right) 
+ 4 \left(\t{B}^2 - A^2 \right) - 2 \left(B^2 - \t{A}^2 \right) 
- 2 \varphi^{\fr23} Q^2 e^{-4H}
\eeal
\label{dilaton-eqn}
\ee
\be
\beal 
\pr_{\rho} \left( \varphi^{-\fr23} r A e^{2H} \right) & = 
2 \varphi^{-\fr13} r B Q e^{D} \;; &\quad
\pr_{\rho} \left( \varphi^{\fr13} B e^{F+2H} \right) & = 
2 \varphi^{\fr23} A Q e^{F+D}\\
\pr_{\rho} \left( \varphi^{\fr13} r \wt{A} e^{2H} \right) & = 
2 \varphi^{\fr23} r \wt{B} Q e^{D} \;; &\quad
\pr_{\rho} \left( \varphi^{-\fr23} \wt{B} e^{F+2H} \right) & =  
2 \varphi^{-\fr13} \wt{A} Q e^{F+D} \\
\eeal
\label{gauge-eqns}
\ee
\be
A\wt{B} + \wt{A}B =0
\label{gaugeconstraint}
\ee
for the 5D dilaton and gauge equations.

This system appears to involve eight unknown functions, but we 
see that in the Lifshitz solutions, one of the two sets of fluxes must
be zero to satisfy \eqref{gaugeconstraint}, and therefore at most we
turn on either the tilded or the untilded fluxes but never both.
Thus, in a given flow we will have six unknown functions. 
These will be subject to seven equations: (\ref{einstein-eqns},
\ref{dilaton-eqn}), and two equations from (\ref{gauge-eqns}). 
As usual, one of the equations in (\ref{einstein-eqns}) is redundant 
because of the Bianchi identity. 

\subsection{AdS$_5$ asymptotic solution}

In the ansatz \eqref{metric}, we have sliced our five dimensional
space-time with two dimensional hyperbolic slices and 
$2+1$ dimensional planar slices. As such therefore, there is no
solution for $F, D$, and $H$ which is globally AdS$_5$, however,
there are solutions which asymptote to AdS$_5$ at large $r$, where
the curvature of the hyperbolic space is effectively suppressed. 
These solutions will have 
\be 
F \sim \rho \;,\quad
D \sim D_0\;,\quad
H \sim H_0 + \rho 
\ee
as $\rho \to \infty$, and will have a constant 5D dilaton, 
$\varphi \sim \varphi_0$, and vanishing  gauge fluxes,
$A \sim B \sim \wt{A} \sim \wt{B}\sim0$ to leading order.
Substituting this in (\ref{einstein-eqns}, \ref{dilaton-eqn},
\ref{gauge-eqns}), the leading order equations fix 
\be
\beal
4 e^{-2D_0} & = \fr16 \left(\varphi_0^{-\fr23} 
+ 2\sqrt{2}\varphi_0^{\fr13}\right), \\
0 & = \fr12 \left(-\varphi_0^{-\fr23} + \sqrt{2}\varphi_0^{\fr13}\right), 
\eeal
\ee
which can easily be solved to find
\be
\varphi_0 = \fr{1}{\sqrt{2}} \quad D_0 = \fr43 \ln{2}.
\label{ads5-solution}
\ee
These asymptotically AdS$_5$ solutions exist for any values 
of  $H_0$ and the topological charge $Q$.

\subsection{AdS$_3 \times \mathcal{H}_2$ solution}

In \cite{Maldacena:2000mw}, a supersymmetric 
AdS$_3 \times \mathcal{H}^2$ solution was considered. 
Here we regard this as part of a one-parameter family of  
AdS$_3 \times \mathcal{H}^2$ solutions in the ansatz 
\eqref{metric}. In appendix A, we consider a more general 
two-parameter family of AdS$_3$ solutions by turning on two fluxes. 

We will get an AdS$_3 \times \mathcal{H}_2$ spacetime from 
the metric (\ref{metric}) by taking constant values for $H=H_0$ 
and $D_0$, and setting $F(\rho) = \rho$. It is easy to check that 
the system has such a solution for constant 5D dilaton field $\varphi_0$ 
and vanishing bulk gauge fluxes $A=\wt{A}=B=\wt{B}=0$ if 
\be
e^{-2D_0} = \fr{\varphi_0^{\fr13}}{2\sqrt{2}}\;, \quad
e^{-2H_0} = \fr{1}{2\varphi_0^{\fr23}}\;,\quad
Q^2 = \varphi_0\sqrt{2} -1. 
\label{AdS3 solution}
\ee
Therefore, we have a family of AdS$_3$ solutions, parametrized 
by the value of 5D dilaton field $\varphi_0$, which should be in the 
range $\varphi_0 \in [ \fr{1}{\sqrt{2}},\infty ) $. These solutions
are illustrated by a grey line in figure \ref{fig:solplot}.

\subsection{Li$_3 \times \mathcal{H}_2$ solution}

We now review the Lifshitz solutions obtained in 
\cite{Gregory:2010gx}. As noted above, such solutions are 
obtained by taking either the tilded or untilded fluxes to vanish. 
The solutions are obtained from our ansatz by setting 
$F(\rho) = z  \rho$, and taking constant functions $H=H_0$ 
and $D = D_0$ as in the AdS$_3$ solutions. 

\subsubsection{Tilded Lifshitz solution $z \geq 1$}

If we turn on a \tit{tilded} pair of gauge fluxes $\wt{A} = \wt{A}_0$, 
$\wt{B} = \wt{B}_0$ for some constant values $\wt{A}_0$ and 
$\wt{B}_0$, ($A = B \equiv 0$) then  (\ref{einstein-eqns},
\ref{dilaton-eqn}, \ref{gauge-eqns}) are satisfied if 
\be
\beal
\varphi_0 & = \fr{\sqrt{2}(z+1)}{2z^2+3z-2}, \quad&
\wt{A}_0^2 & = \fr{z(z-1)}{2} e^{-2D_0}, \\
e^{-2D_0} & = \left[2(z+1)^2(2z^2+3z-2)\right]^{-\fr13}, \quad&
\wt{B}_0^2 & = \fr{z-1}{2} e^{-2D_0}, \\
e^{-2 H_0} & = \fr32 z e^{-2D_0}, \quad&
Q^2 & = \fr{2z^2 + 3z-2}{9z}. 
\eeal
\ee
This family of solutions is parametrized by the value of the 
dynamical exponent $z$, which in this case should be greater 
than one, and is shown in figure \ref{fig:solplot} as a blue
line. 

\subsubsection{Untilded Lifshitz solution $1 \leq z \leq 2$}

\begin{figure}[ht]
\centering
\includegraphics[width=0.7\textwidth]{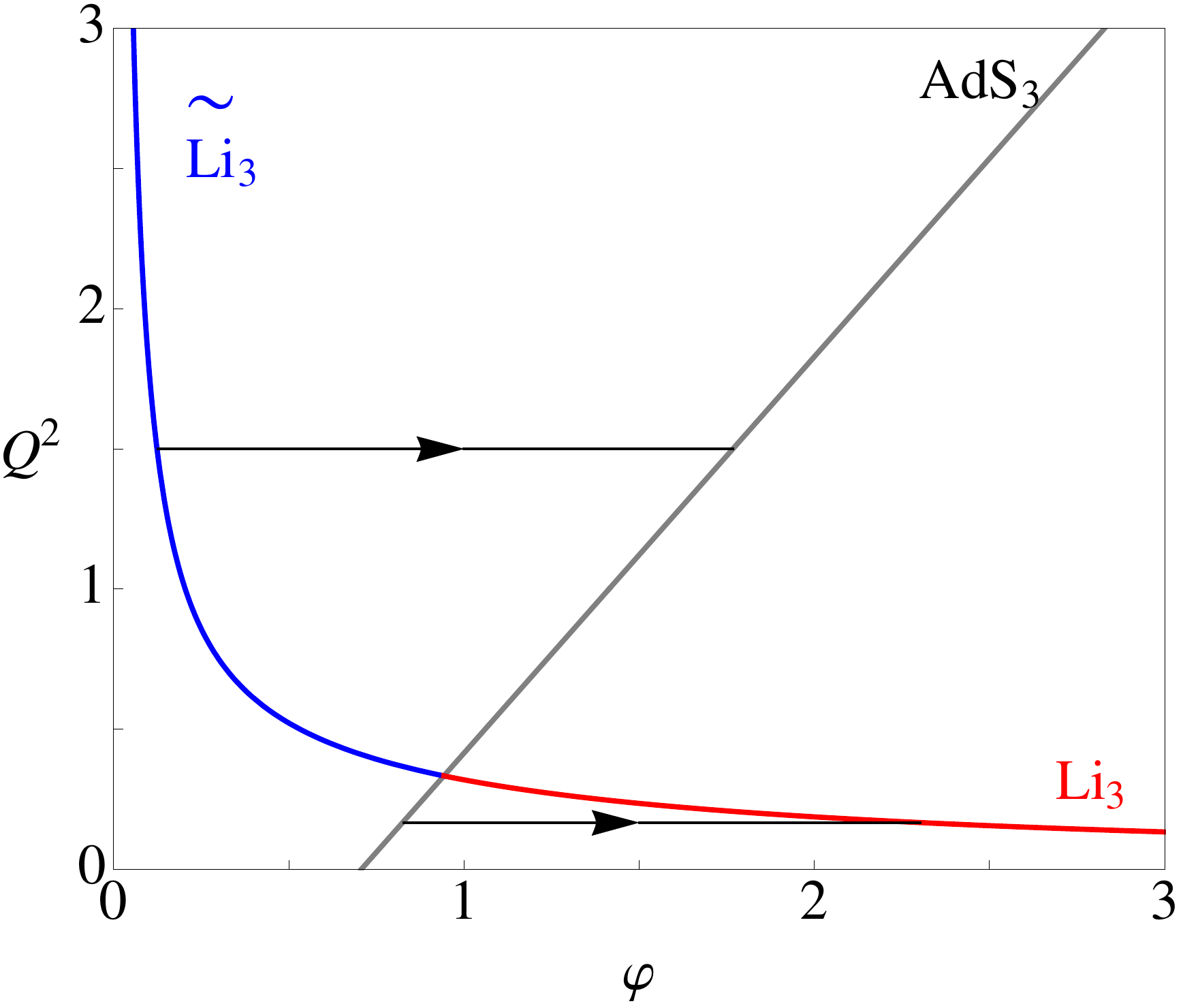}
\caption{The values of $Q, \varphi_0$ for the AdS$_3$, 
$\wt{Li_3}$ and Li$_3$ solutions. The AdS$_3$ family is 
parametrized by $\varphi_0$, which determines $Q^2 
= \sqrt{2} \varphi_0 -1$. The Lifshitz families are parametrized 
by $z$, which determines $Q$ and $\varphi_0$. Also shown
are flows between the solutions, which must occur at constant
$Q$, with an arrow depicting the direction of the flow.} 
\label{fig:solplot}
\end{figure}

If we turn on the other pair of fluxes, i.e.\ \tit{untilded} gauge 
fluxes $A = A_0$, $B = B_0$ for some constant values $A_0$ 
and $B_0$, ($\wt{A} = \wt{B} \equiv 0$) then  (\ref{einstein-eqns},
\ref{dilaton-eqn}, \ref{gauge-eqns}) are satisfied if 
\be
\beal
\varphi_0 & = \fr{\sqrt{2}z(z+1)}{-2z^2+3z+2}, \quad&
A_0^2 & = \fr{z(z-1)}{2} e^{-2D_0},  \\
e^{-2D_0} & = \left[2z^2(z+1)^2(-2z^2+3z+2)\right]^{-\fr13}, \quad&
B_0^2 & = \fr{z-1}{2} e^{-2D_0},  \\
e^{-2 H_0} & = \fr32 z e^{-2D_0}, \quad&
Q^2 & = \fr{-2z^2 + 3z+2}{9z}. 
\eeal
\ee
This second family of solutions is again parametrized by $z$, but this 
must now lie in the range $1 \leq z \leq 2$ which gives positive $Q^2$. 
These solutions are shown as a red line in the $\left(Q^2,\varphi_0\right)$ 
plane in Figure~\ref{fig:solplot}. 

\section{RG flow solutions}
\label{flows}

We now turn to the construction of flows interpolating between 
the solutions reviewed in the previous section. Such interpolating 
solutions correspond to RG flows in the dual field theory, with the 
solution at small $r$ corresponding to the IR limit of the RG flow, 
and the solution at large $r$ corresponding the the UV limit 
of the RG flow. The study of such holographic flows was 
initiated in \cite{Girardello:1998pd,Freedman:1999gp}.

Analogous flows were previously constructed for the Type IIA 
theory in \cite{Braviner:2011kz}. As in that case, the charge 
$Q$ will be conserved along the flows; flows will move 
horizontally in figure~\ref{fig:solplot}. Therefore the solutions 
that can be related by flows are the $\wt{{\rm Li}}_3$ and AdS$_3$ 
for large enough values of  $Q$, and AdS$_3$ and Li$_3$ for 
smaller values of  $Q$. There is also the possibility of having 
flows which start from  the asymptotically AdS$_5$ solution in 
the UV, which exists for any value of the charge $Q$, and 
approach any of these AdS$_3$ or Lifshitz solutions in the IR. 

\subsection{Linearized analysis}

Before we proceed to the construction of the actual flows, 
we will perform a linearized perturbation analysis around 
each of the fixed-point solutions, to determine which 
direction we would expect the flows to go in (that is, 
which solution should be in the IR and which in the UV). 
This corresponds to computing the dimensions of the 
deforming operators in the dual field theories. We then
construct the interpolating solutions numerically.  

\subsubsection{Linearisation around AdS$_5$}
\label{ads5exp}

The expansion around the asymptotically AdS$_5$ solution is a 
little more conceptually involved than the others, because 
AdS$_5$ is not an exact solution of the equations of motion, 
but only an asymptotic solution. We can avoid these subtleties 
by imagining that we take the radius of curvature of the compact 
hyperbolic space to zero by taking $h_0 \to \infty$, and neglecting 
terms in the equations of motion involving $e^{-2h_0}$. This 
will give us the linearized form of the equations of motion 
around the pure AdS$_5$ solution which will allow us to 
read off the scaling of the linearized solutions. These scalings 
will remain valid for the linearized modes in the asymptotically 
AdS$_5$ solution with finite $h_0$ to leading order at large 
$r$, as the physical volume of the compact hyperbolic 
space diverges as $r \to \infty$. 

We write the solution as
\be
\beal
\pr_{\rho}F & = 1 + y_0(\rho), \quad
& D & = D_0 + y_1(\rho), \quad
& A & = y_8(\rho), \\
H & = \rho + H_0 + y_2(\rho), \quad
& \pr_{\rho} H & = 1 + y_4(\rho), \quad 
& B & = y_9(\rho), \\
\varphi & = \varphi_0 + y_3(\rho), \quad
& \pr_{\rho} \varphi & = 0 + y_5(\rho), 
\eeal
\ee
and linearize in the $y_i$, taking $H_0 \to \infty$. 
At linear order we will not see the constraint 
\eqref{gaugeconstraint}, but we recall that we 
will only consider solutions with either  $\left(y_6,y_7\right)$ 
or $\left(y_8,y_9\right)$, but not all four at the same time.  
The other equations in  (\ref{einstein-eqns}, \ref{dilaton-eqn},
\ref{gauge-eqns}) then give us a system of first-order equations,  
\be
\beal
\dot{y}_0 & = -4y_0, \quad
&\dot{y}_1 &= y_0 - 8y_1 + 2y_4 ,  \quad
&\dot{y}_2 &= y_4,   \\
\dot{y}_3 &= y_5,  \quad
&\dot{y}_4 &= -4y_4,  \quad
&\dot{y}_5 &= -4y_3 - 4y_5, \\
\dot{y}_6  &= -3y_6,  \quad
&\dot{y}_7 &= -3y_7,  \qquad
\dot{y}_8 = -3y_8, \quad
&\dot{y}_9 &= -3y_9, 
\eeal
\label{lin-sys-AdS5}
\ee
and a constraint equation, 
\be \label{ceq}
y_1 = \frac{y_0 + 2 y_4}{4}. 
\ee
We can easily verify that this constraint is consistent with the 
first-order system. Imposing the constraint, and keeping one 
of the two pairs of gauge fluxes, we will have a seven-dimensional 
space of linearized solutions.  For example, for the case where 
we keep $\left(y_8,y_9 \right)$, the linearized solutions are 
\be
\beal
\partial_\rho F & = 1 + C_0 e^{-4\rho}, \quad&
\varphi & = \varphi_0 + \l \rho e^{-2\rho} + \eta e^{-2\rho}, \\
D & = D_0  + \frac{1}{4} (C_0 + 2 C_4) e^{-4 \rho},  \quad&
A & =  C_8 e^{-3\rho},  \\
H & = \rho + H_0 + C_2 - \frac{1}{4} C_4 e^{-4 \rho},  \quad&
B &=  C_9 e^{-3\rho}.  
\eeal
\ee
These solutions correspond to infinitesimal VEVs and sources 
for corresponding operators. The constants $C_0, C_4$ are 
the energy density and an anisotropic pressure; the corresponding 
sources are deformations of the boundary metric. These are 
$C_2$ and a constant $F_0$ in $F$, which we can freely add 
since the equations of motion only involve $\partial_\rho F$. 
Both $C_2$ and $F_0$ are pure gauge degrees of freedom; 
the former corresponds to shifting the background $H_0$, 
and the latter is a pure diffeomorphism. The parameters 
$C_8$ and $C_9$ are charge densities for the gauge fields; 
the corresponding sources are constant components of the 
vector potentials, which are pure gauge, and are also absent 
from our ansatz since we wrote it in terms of the field strengths. 
Finally $\l$ and $\eta$ are the source and VEV for the operator 
corresponding to the 5D dilaton. This operator is particularly 
interesting to us as we will see that the flows from AdS$_5$ 
to the AdS$_3$ and Lifshitz solutions will involve turning on 
this source. As this is a relevant deformation, we would 
expect flows from AdS$_5$ in the UV, approaching the 
other solutions in the IR. 

Since they do not enter into the equations of motion in our 
ansatz, the constant part of $F$ and the constant part of the 
gauge potentials will not play any role in the flows we consider. 
This is a remarkable fact; it implies that in the flows from 
AdS$_5$ to Lifshitz, the only physical source we can find 
turned on at the AdS$_5$ end of the flow is $\l$. This does 
not break the Lorentz invariance. Thus, when we have a 
flow to Lifshitz, the breaking of the Lorentz invariance along 
the flow is spontaneous. 

\subsubsection{Linearisation around AdS$_3$ solutions}

\begin{figure}
\centering
\includegraphics[width=0.4\textwidth]{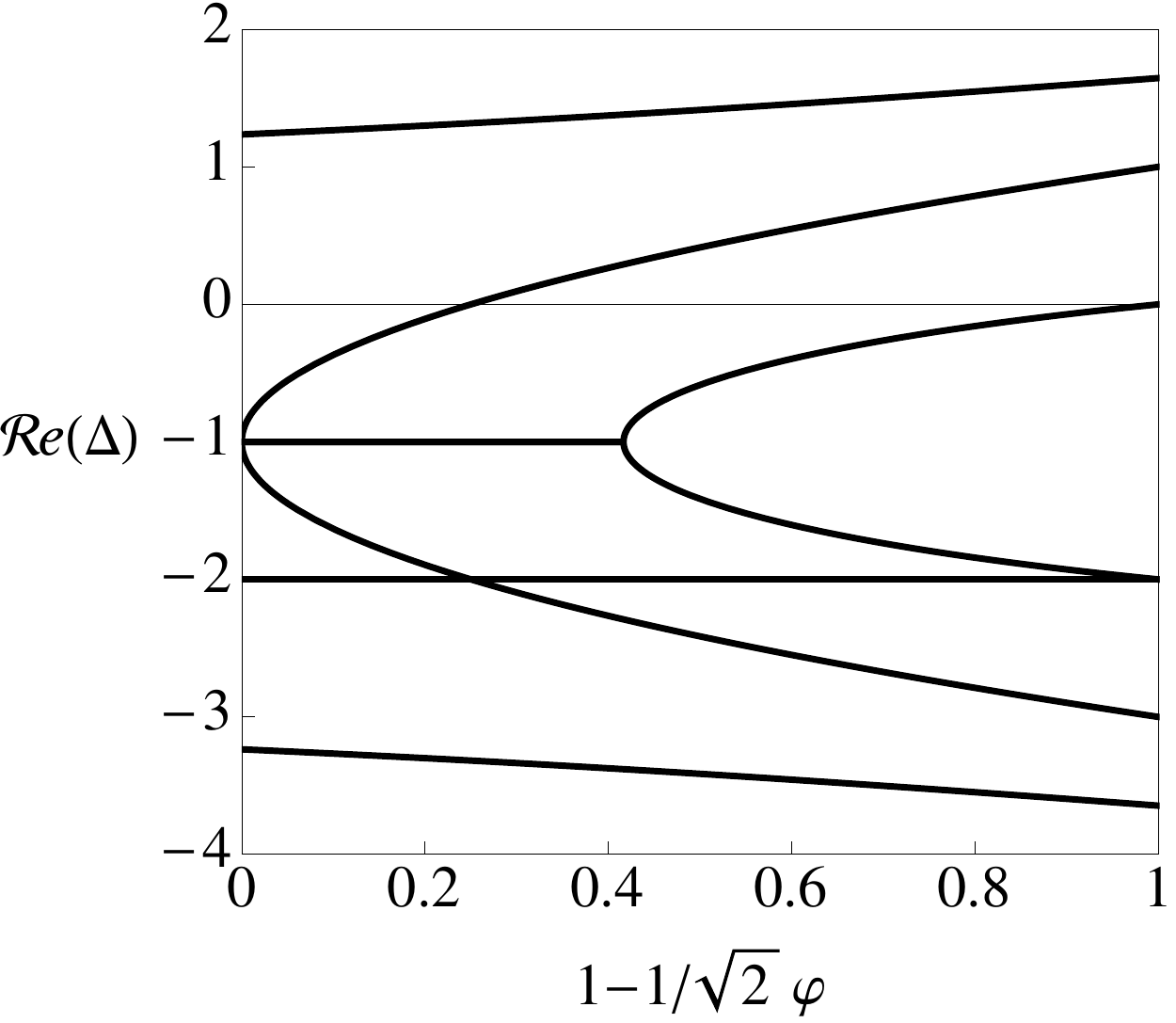}~
\includegraphics[width=0.4\textwidth]{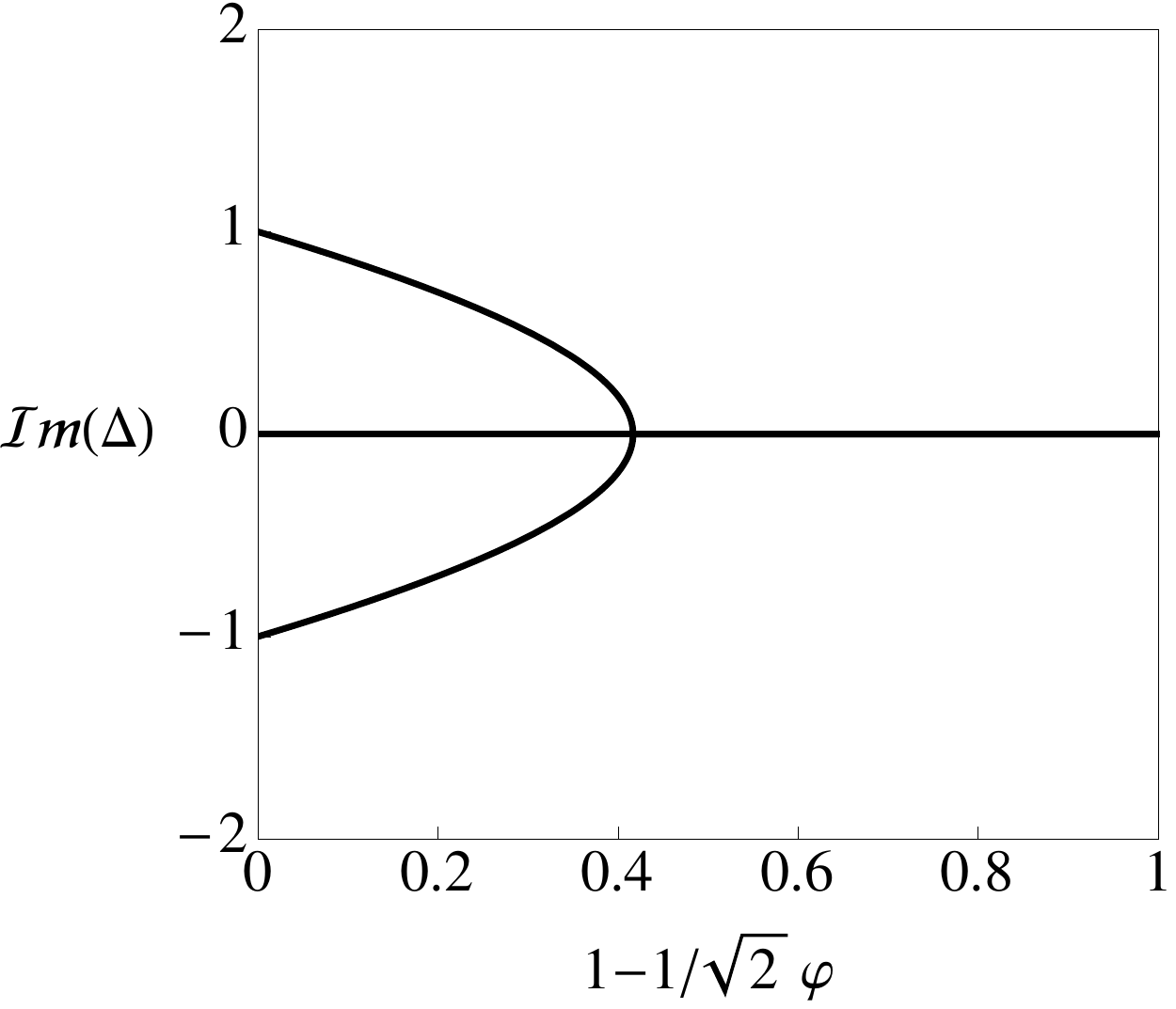}
\caption{Plots of real and imaginary parts of the eigenvalues 
of the linear perturbations from the AdS$_3$ solution as 
functions of the background value of the 5D dilaton field $\varphi_0$.} 
\label{eigen-ads3}
\end{figure}

We expect to have flows relating AdS$_3$ to both $\wt{{\rm Li}}_3$ 
and Li$_3$ spacetimes, therefore it is interesting to consider 
perturbations for both tilded and untilded fluxes in this case. 
Hence, we have the following linear perturbation from 
the AdS$_3$ solution
\be
\tbf{X} = \tbf{X}_0 + \tbf{y},
\ee
where $\tbf{X}_0 = \left(F',D,H,\varphi,H',\varphi',\wt{A},\wt{B},A,B\right) 
= \left(1,D_0,H_0,\varphi_0,0,0,0,0,0,0\right)$ is the fixed point 
solution corresponding to the AdS$_3 \times \mathcal{H}_2$ 
spacetime and $\tbf{y}(\rho)$ is a vector of perturbations. 
Linearising the equations of motion around the fixed point 
gives us a linear system
\be
\dot{\tbf{y}} = \mathbb{A}_{AdS_3} \cdot \tbf{y},
\ee
together with a constraint equation analogous to \eqref{ceq}. 
The matrix $\mathbb{A}_{AdS_3}$ is a $10\times10$ matrix
dependent on the background field values, however,
as with the AdS$_5$ case, we may only switch on either the
tilded or untilded fluxes, which both have exactly the same 
form of perturbation equations. In addition, the Bianchi identity
implies a zero mode, thus our effective perturbations are 
reduced to a seven-dimensional system 
\be
\dot{\tbf{y}}_{\rm{red}} = \mathbb{A}_{\rm{red}} \cdot \tbf{y}_{\rm{red}},
\label{lin-sys-AdS3}
\ee
where ${\tbf{y}}_{\rm{red}}=
(\delta F',\delta H,\delta \varphi,\delta H',\delta \varphi',
\delta A(\delta\wt{A}),\delta B(\delta\wt{B}))$, and
writing $c = \sqrt{2}/\varphi_0$:
\be
\mathbb{A}_{\rm{red}}  = \left [
\begin{matrix}
-2 & 0 & 0 & 0 & 0 & 0 & 0 \\
0 & 0 & 0 & 1 & 0 & 0 & 0 \\
0 & 0 & 0 & 0 & 1 & 0 & 0 \\
0 & \textstyle{\frac{16-2c}{3}} 
& \textstyle{\frac{\sqrt{2}c}{9}(c-2)} & -2 & 0 & 0 & 0 \\
0 & \textstyle{\frac{4\sqrt{2}}{c}(c-2)} 
& \textstyle{\frac{2-4c}{3}} & 0 & -2 & 0 & 0 \\
0 & 0 & 0 & 0 & 0 & -1 & \sqrt{4-2 c} \\
0 & 0 & 0 & 0 & 0 & \sqrt{4-2 c} & -1
\end{matrix}
\right ]
\ee
In this format we see the perturbation of the flux decouples from
the geometry, and the equation for $\delta F'$ also decouples.
This matrix has a set of eigenvalues $\left\{\Delta_i\right\}$,
\be
\Delta_i = -2\,; \quad -1 \pm \sqrt{4 - c \pm \sqrt{9 - 2c+ c^2}}\,;
\quad -1 \pm \sqrt{4-2c}\;,
\ee
with corresponding eigenvectors $\left\{\tbf{v}_i\right\}$, thus 
the solution of the linear system (\ref{lin-sys-AdS3}) is
\be
\tbf{y}_{\rm{red}} = \sum_i \tbf{v}_i e^{\Delta_i \rho}.
\ee
The eigenvalues are plotted in figure 
\ref{eigen-ads3}, and we see that as in \cite{Braviner:2011kz}, 
some of the eigenvalues are complex for some values of 
$\varphi_0$, signalling a potential instability of these solutions. 
We will return to this issue at the end of our analysis.

Clearly, the $\Delta = -2$ eigenvalue corresponds to a pure geometry
fluctuation, and actually corresponds to the fluctuation from a mass. 
The final pair of eigenvalues 
$\Delta_{\pm} = -1 \pm \sqrt{4-\fr{2\sqrt{2}}{\varphi_0}}$
switch on flux, 
hence corresponding operators on the field theory side are 
relevant when $\Delta_{+}<0$, 
i.e.\ for $\fr{1}{\sqrt{2}} < \varphi_0 < \fr{2\sqrt{2}}{3}$. 

Note that  $\varphi_0 =  \fr{2\sqrt{2}}{3}$ corresponds exactly 
to the point where all AdS$_3$, $\wt{{\rm Li}}_3$ and Li$_3$ solutions 
coincide. Hence, for $\fr{1}{\sqrt{2}} < \varphi_0 < \fr{2\sqrt{2}}{3}$ 
we will have a relevant operator near AdS$_3$.  If we excite 
the untilded fluxes, we can then expect a flow from the AdS$_3$ 
solution in the UV to the Li$_3$ solution in the IR. For 
$\varphi_0 > \fr{2\sqrt{2}}{3}$ we will have an irrelevant 
operator near AdS$_3$. So if we excite the tilded fluxes,  
we can expect to have flows from the $\wt{{\rm Li}}_3$ spacetime 
in the UV to the AdS$_3$ spacetime in IR. These expected 
flows are  presented in Figure~\ref{fig:solplot}. We 
will construct these flows numerically below. 

In addition to the flux deformations, we see from figure 
\ref{eigen-ads3} that there is one deformation which is always 
irrelevant. This should correspond to the flow approaching 
AdS$_3$ from the asymptotically AdS$_5$ solution. 

\subsubsection{Linearisation around $\wt{{\rm Li}}_3$ solutions}

\begin{figure}
\centering
\includegraphics[width=0.4\textwidth]{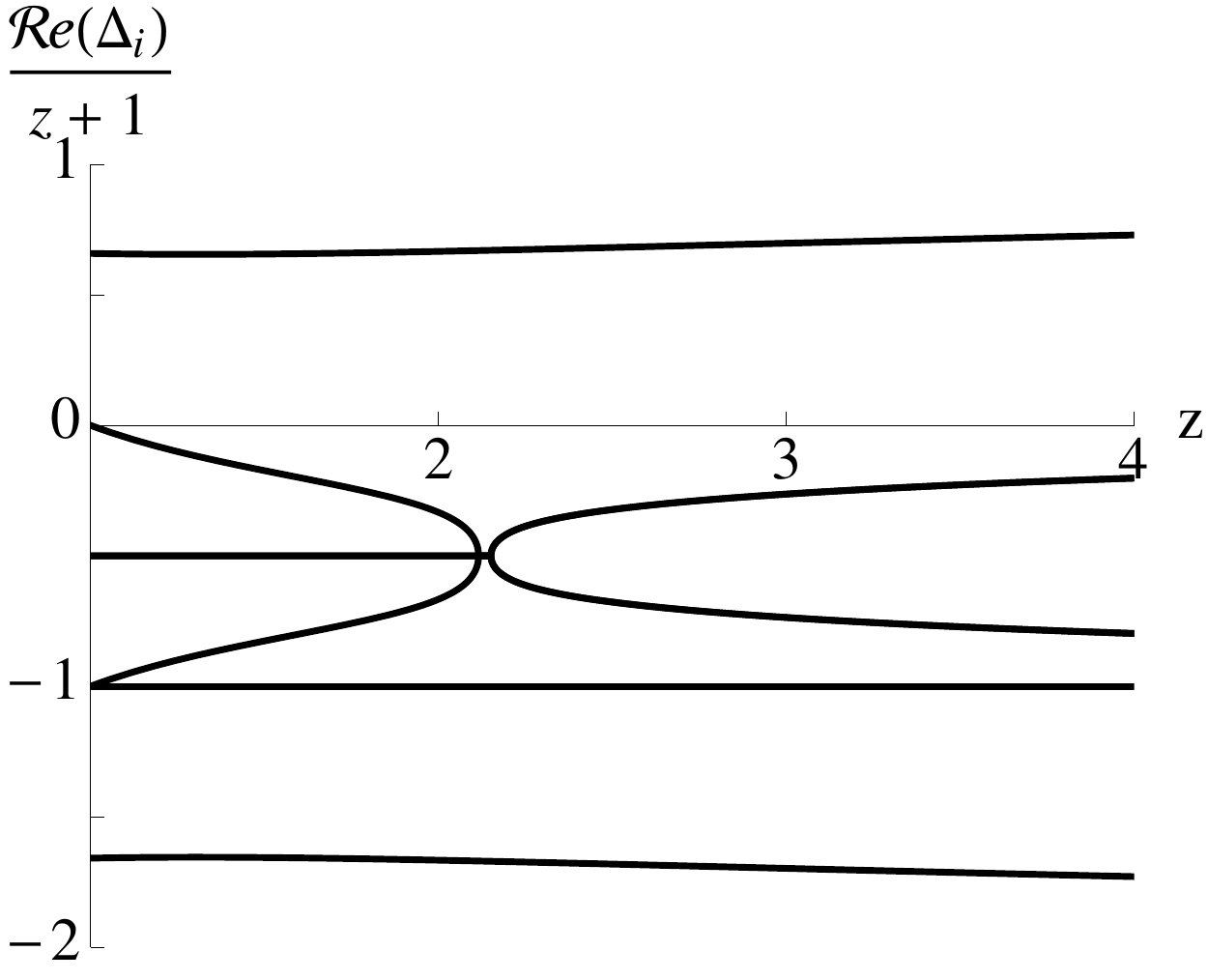}~
\includegraphics[width=0.4\textwidth]{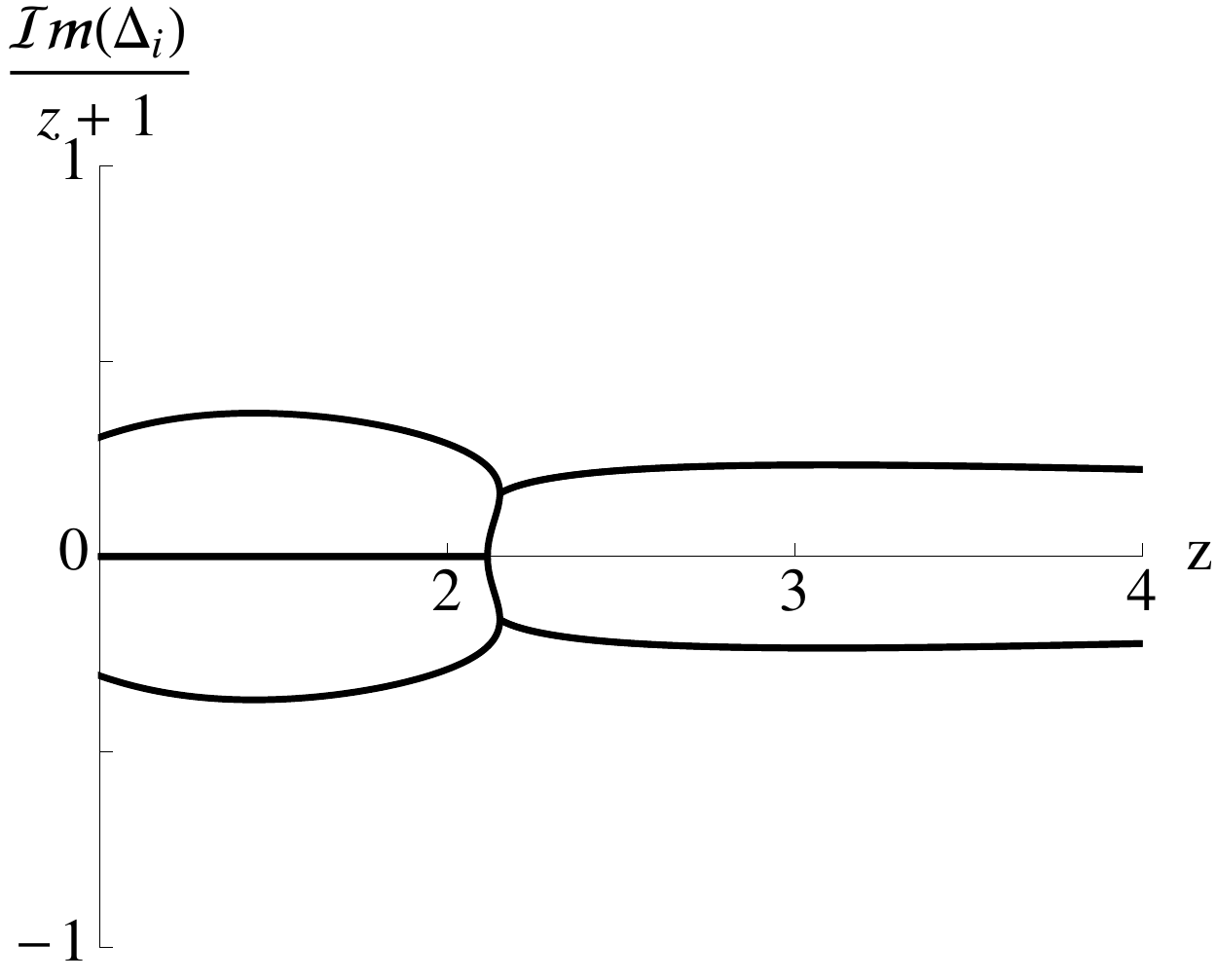}
\caption{Plots of the real and imaginary parts of the eigenvalues 
of the linear perturbations from the $\wt{{\rm Li}}_3$ solutions, 
divided by $z+1$, as functions of the background values of 
the dynamical exponent $z$.} 
\label{eigen-wtli3}
\end{figure}

In this case we must set the untilded fluxes to zero identically 
to satisfy \eqref{gaugeconstraint}. 
We write the variables as 
\be
\tbf{X} = \tbf{X}_0 + \tbf{y},
\ee
where $\tbf{X}_0 = \left(F',D,H,\varphi,H',\varphi',\wt{A},\wt{B}\right) 
= \left(z,D_0,H_0,\varphi_0,0,0,\wt{A}_0,\wt{B}_0\right)$ 
are the background values and $\tbf{y}$ are the linear 
perturbations. This gives a linear system 
\be
\dot{\tbf{y}} = \mathbb{A}_{\wt{Li}_3} \cdot \tbf{y}
\label{lin-sys-wtLi3}
\ee
together with a constraint equation analogous to \eqref{ceq}. 
The entries of the matrix $\mathbb{A}_{\wt{Li}_3}$ are 
parametrized by the value of dynamical exponent $z$, and 
although the corresponding eigenvalues can be found analytically
(in terms of square roots of solutions to a cubic) their form is not particularly
illuminating thus we present them only graphically in figure
\ref{eigen-wtli3}. The eigenvalues occur in pairs
with the sum of each pair equal to $-(z+1)$. We see that we 
have complex eigenvalues for all values of $z$ along this family. 
We also note that there is a single irrelevant mode, 
corresponding to the expected flow approaching this solution 
from the asymptotically AdS$_5$ solution. 

\subsubsection{Linearisation around Li$_3$ solutions}

\begin{figure}
\begin{center}
\includegraphics[width=0.45\textwidth]{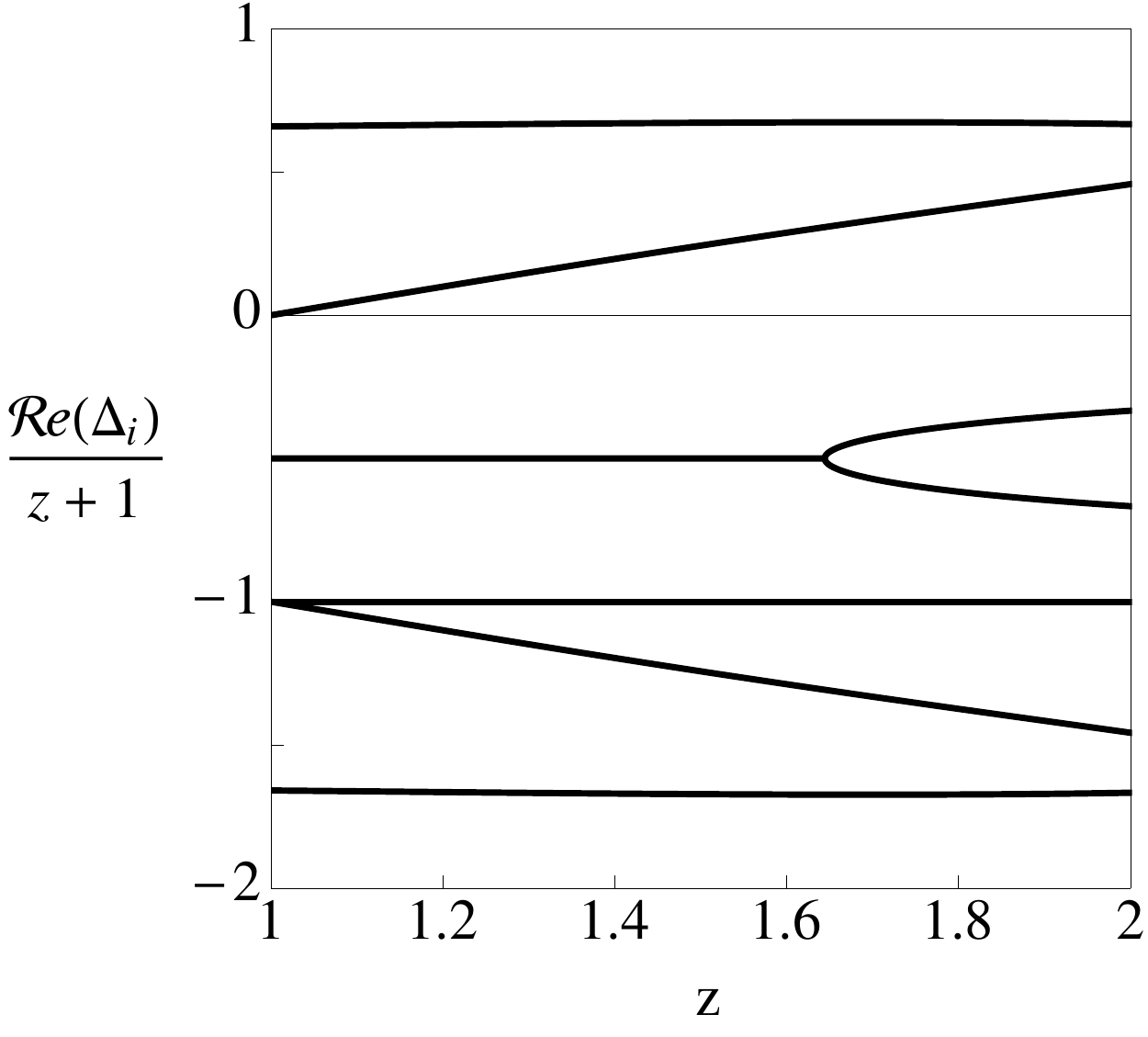}~~
\includegraphics[width=0.45\textwidth]{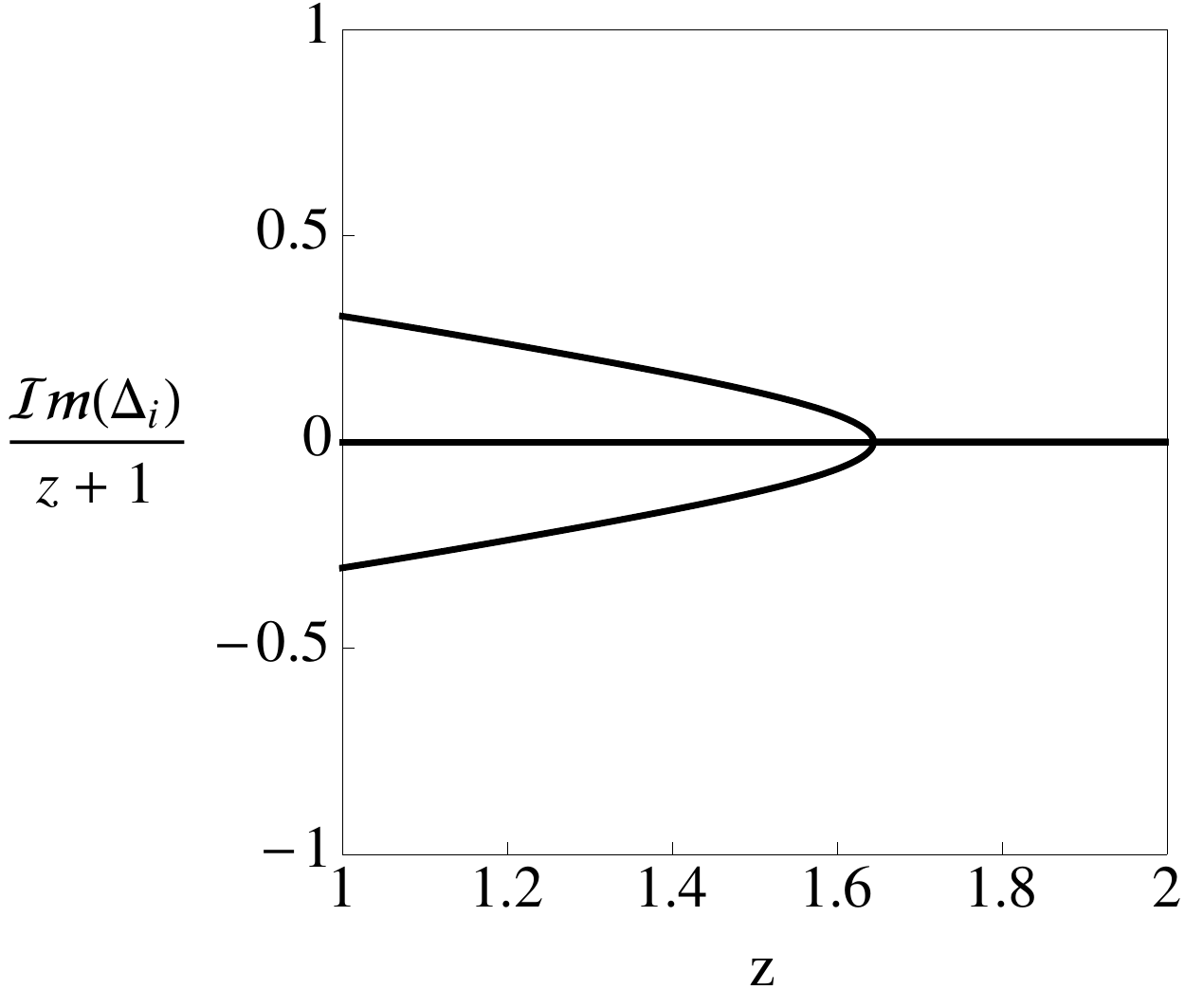}
\caption{Plots of the real and imaginary parts of the 
eigenvalues of linear perturbations from the Li$_3$ 
solutions, divided by $z+1$, as functions of the background 
values of the dynamical exponent $z$, in this case $1 \leq z \leq 2$.} 
\label{eigen-li3}
\end{center}
\end{figure}

This is similar to the previous case, although now it is the 
tilded fluxes which must be set equal to zero. We again 
have an 8-dimensional system of linear perturbations, 
with background values  
$\tbf{X}_0  = \left(F',D,H,\varphi,H',\varphi',A,B\right)   
= \left(z,D_0,H_0,\varphi_0,0,0,A_0,B_0\right) $, and a 
linear system with a  matrix $\mathbb{A}_{Li_3}$ and a 
constraint. We will again have seven linearly independent 
modes, with eigenvalues coming in pairs, with the sum of 
the eigenvalues in each pair equal to $-(z+1)$. The 
resulting eigenvalues are presented in figure \ref{eigen-li3}. 
Here we see complex eigenvalues for a range of values of 
$z$ near 1, but there is a range near 2 where all the 
eigenvalues are real and the solutions may be stable. 
We also note that there are two irrelevant modes, 
corresponding to the expected flows approaching this 
solution from asymptotically AdS$_5$ and AdS$_3$ solutions. 

\subsection{Numerical Flows}

Here we present the result of numerical solutions of the full 
non-linear system of equations of motion for the interpolating 
solutions between different fixed points in UV 
$(r \rightarrow \infty)$ and IR $(r \rightarrow 0)$. 
We discuss first the  flows between AdS$_3$ and Li$_3$ 
spacetimes and then consider the flows from the 
asymptotically AdS$_5$ solution in the UV.

\subsubsection{Flows between AdS$_3$ and Li$_3$ spacetimes}

From the linearized analysis, we expect flows from AdS$_3$ 
in the UV to Li$_3$ in the IR and flows from $\wt{{\rm Li}}_3$ in 
the UV to AdS$_3$ in the IR, as depicted in figure 
\ref{fig:solplot}. We constructed examples of these 
flows numerically, using a shooting method. The shooting 
is carried out starting from the IR fixed point at small $r$, 
integrating numerically  to larger $r$. Shooting is required 
to obtain the flows between AdS$_3$ and Li$_3$ because 
the IR fixed point always has two positive eigenvalues, and 
the generic flow will go to the asymptotically AdS$_5$ solution. 
Hence possible directions of shooting lie in the plane  
spanned by the two corresponding  unstable directions 
and can be parametrized by the single angle variable, 
say, $\zeta$.  We find the value of $\zeta$ giving the 
desired flow by bisection of an initial interval of values of $\zeta$. 

\medskip

\noindent $\bullet\quad Q^2 \in \left[0,\fr13\right]$: 
\tit{Flows from AdS$_3$ to Li$_3$}

\medskip

\begin{figure}
\begin{center}
\includegraphics[scale=0.38]{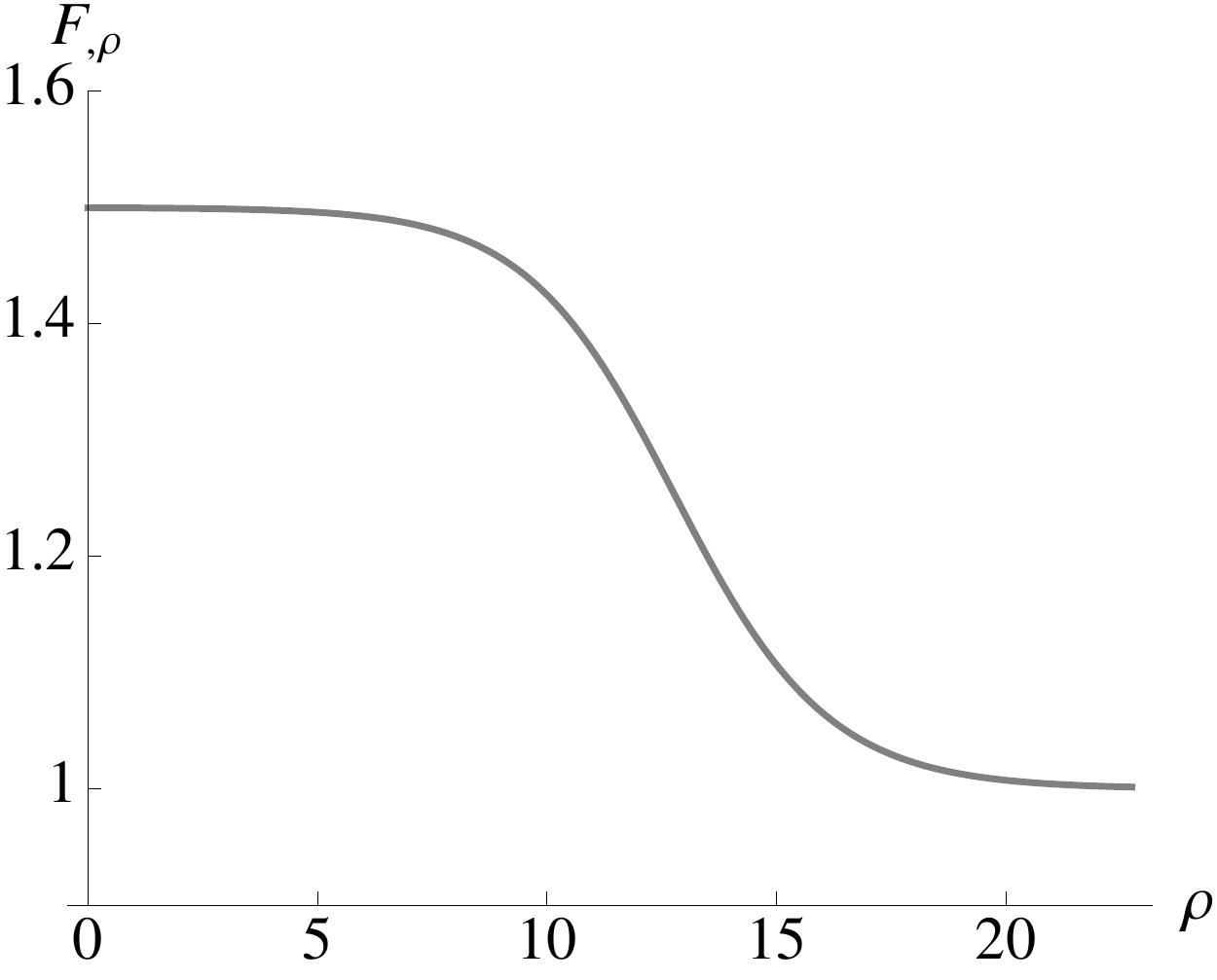}
\includegraphics[scale=0.38]{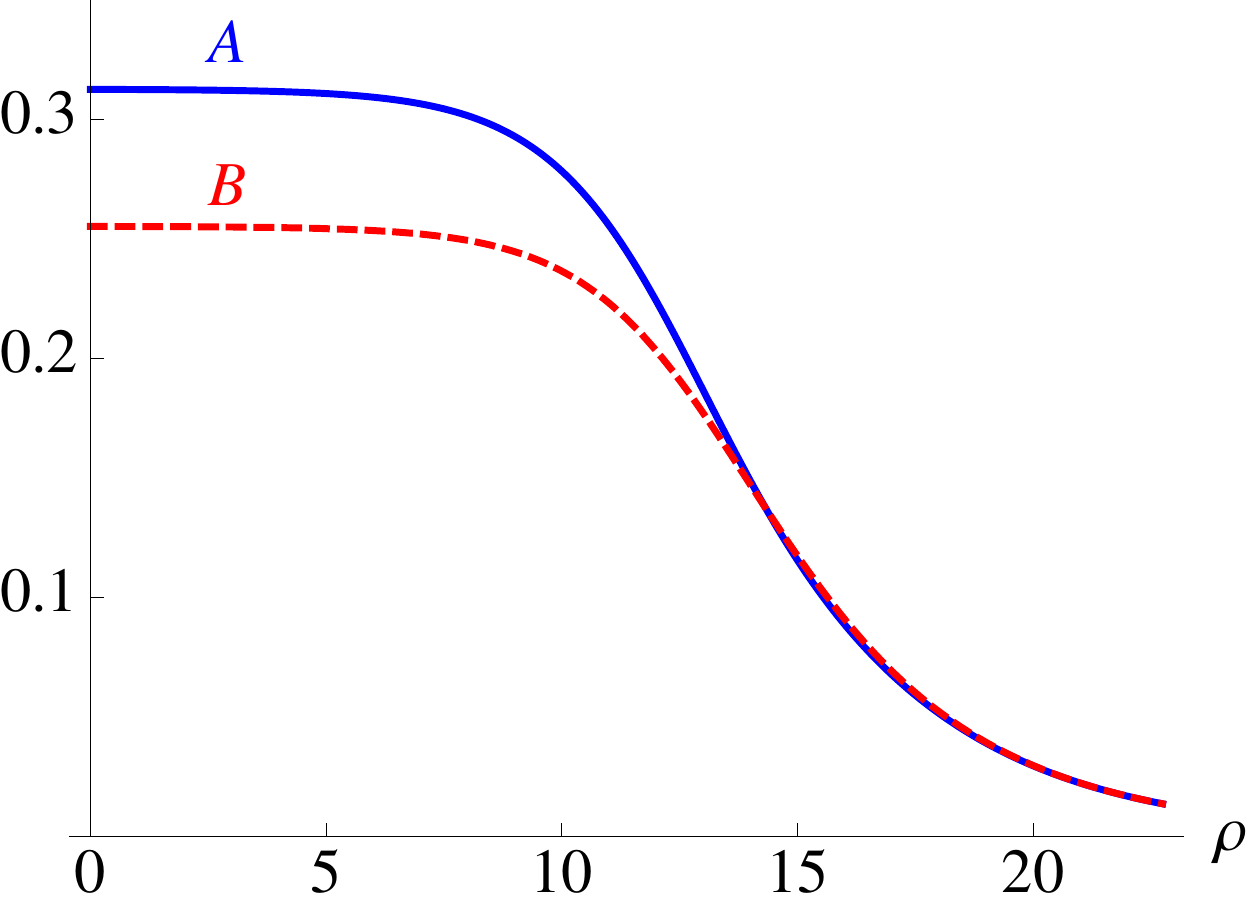}
\includegraphics[scale=0.38]{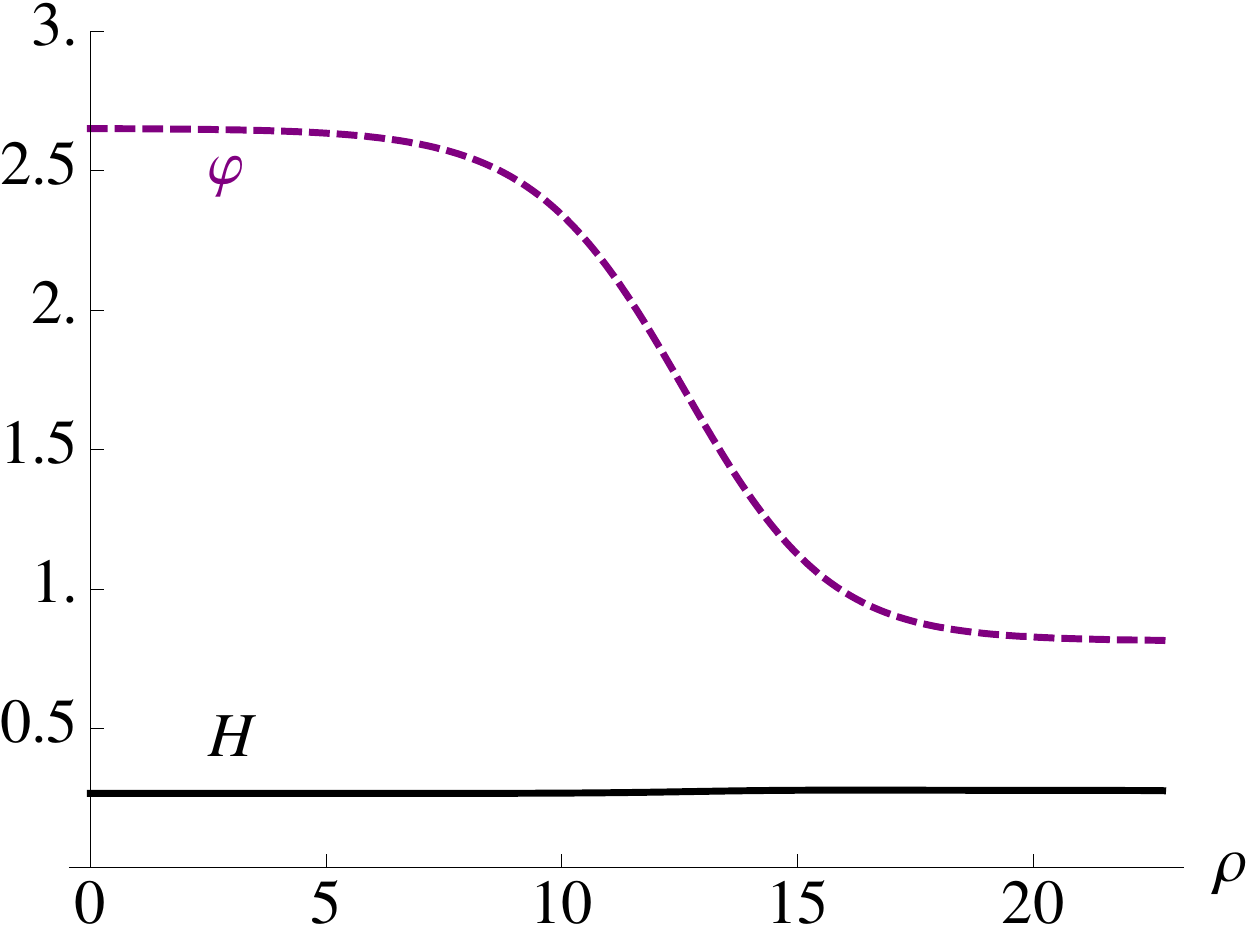}
\caption{Solution interpolating between Li$_3$ with $z=3/2$ 
and AdS$_3$, with $Q^2=\fr{4}{27}$.} 
\label{flows-ads3-li3}
\end{center}
\end{figure}

We present an example of such a solution in figure
\ref{flows-ads3-li3}: this case interpolates between the 
untilded Lifshitz solution with $z=3/2$ for small $r$ (IR) 
and the AdS$_3$ solution for large $r$ (UV) . The plot 
of $F'$ shows that it starts from the value $3/2$ 
and goes to $1$, the other plots show how fluxes of the 
gauge fields go to zero at large $r$.

\medskip

\noindent $\bullet\quad Q^2 > \fr13$: 
\tit{Flows from $\wt{{Li}}_3$ to AdS$_3$}

\medskip

\begin{figure} [ht]
\begin{center}
\includegraphics[scale=0.38]{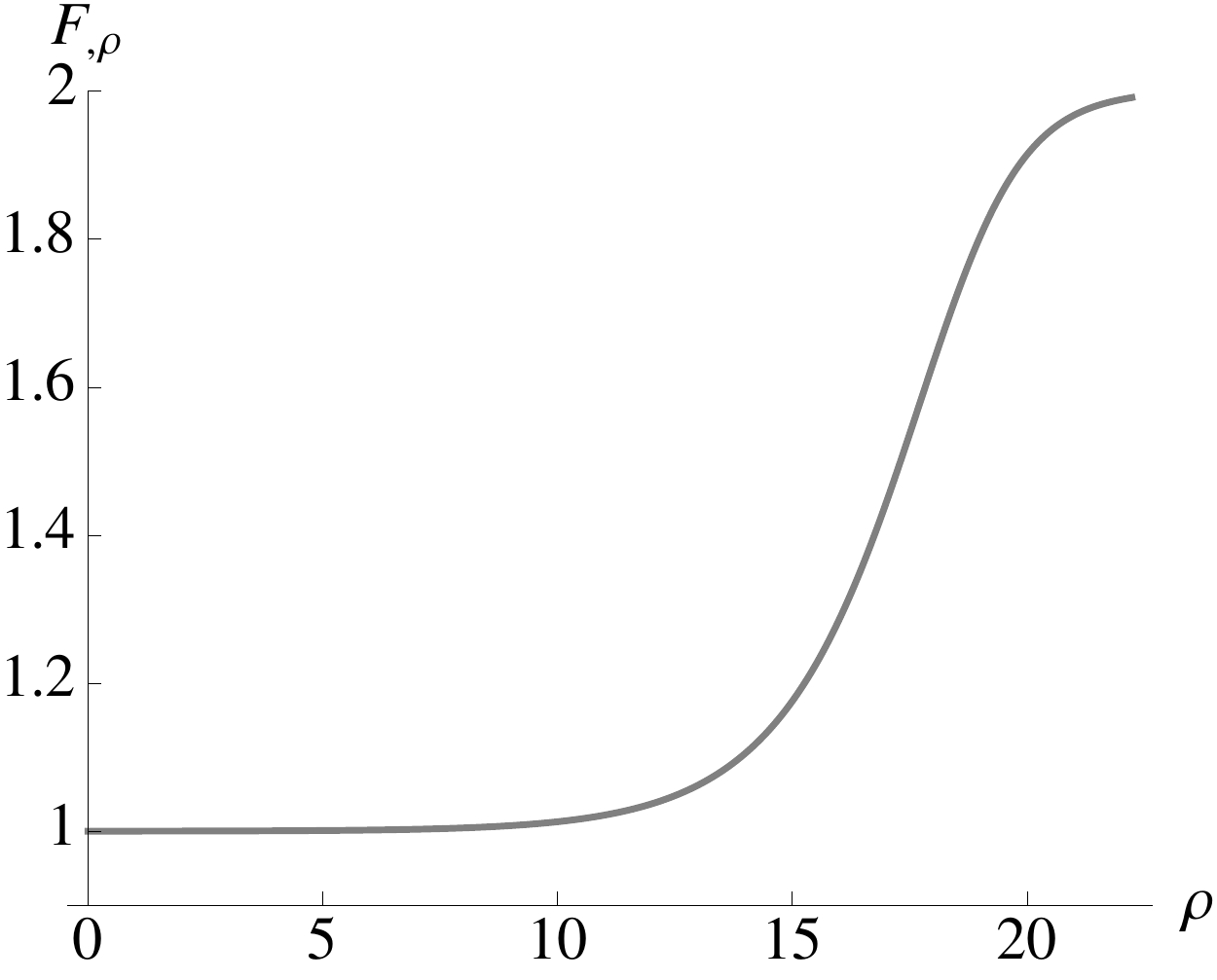}~
\includegraphics[scale=0.38]{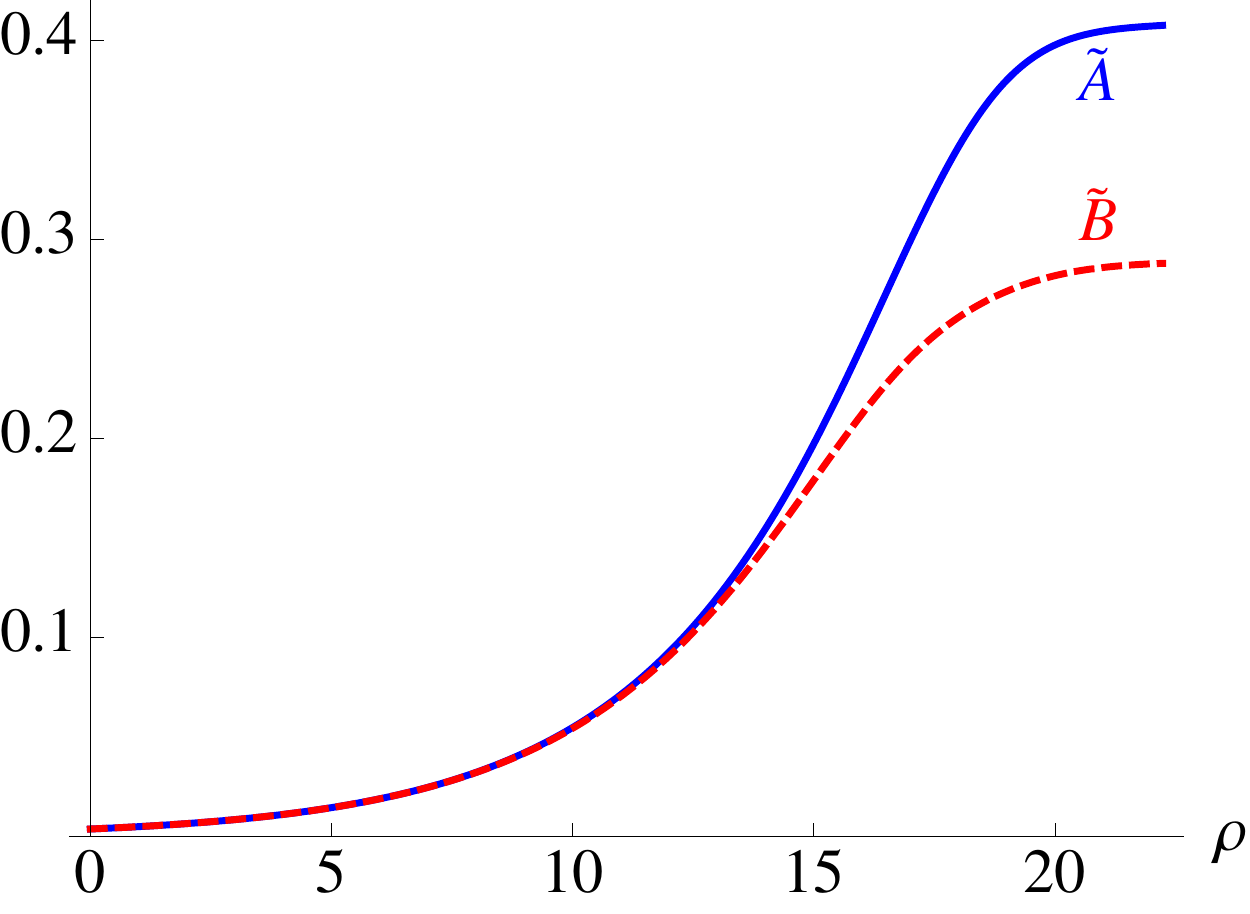}~
\includegraphics[scale=0.38]{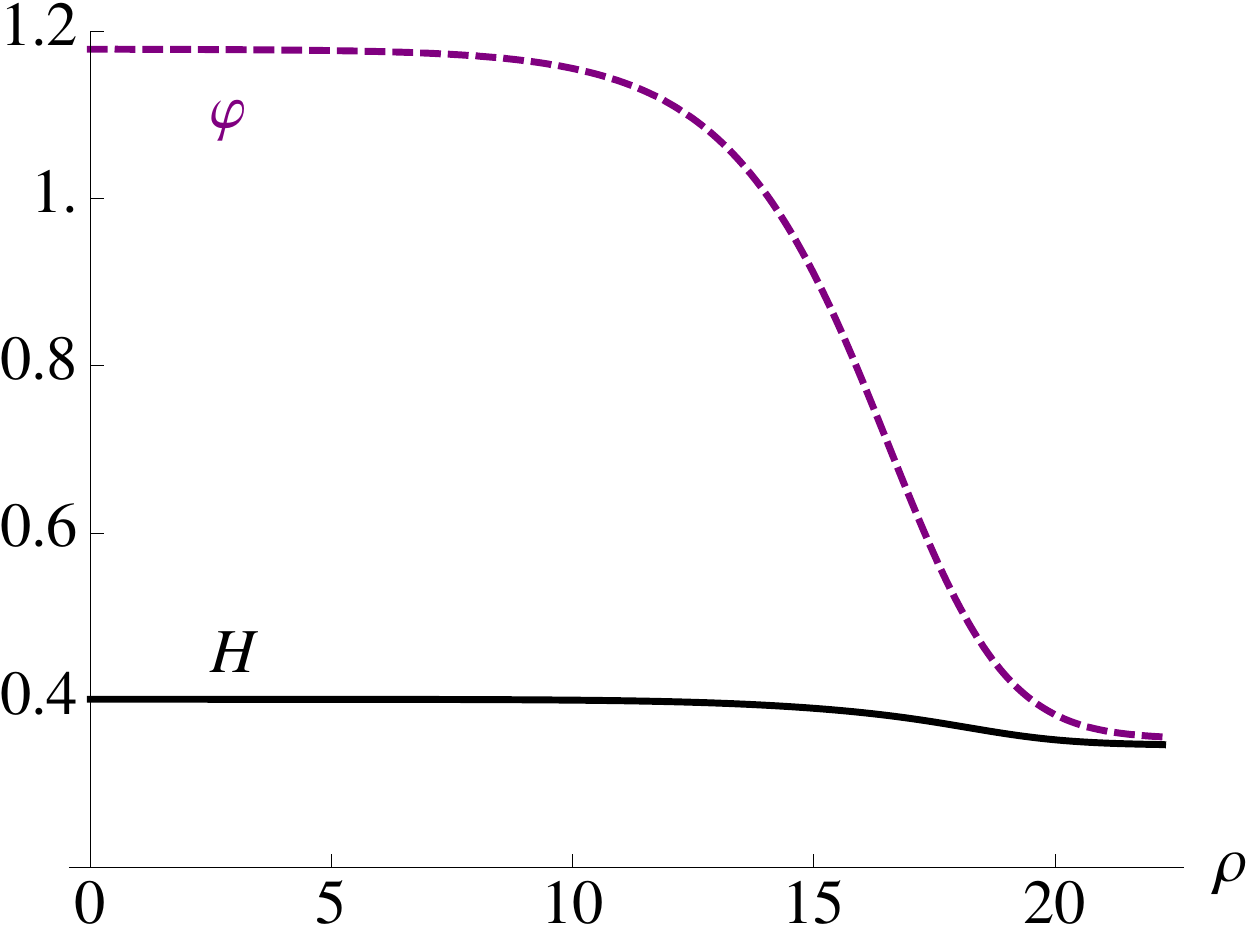}
\caption{Solution interpolating between AdS$_3$ and 
$\wt{{\rm Li}}_3$ with $z=2$, with $Q^2=\fr23$.}
\label{flows-wtli3-ads3}
\end{center}
\end{figure}

We present an example of such a solution in figure
\ref{flows-wtli3-ads3}: this case interpolates between AdS$_3$ 
for small $r$ (IR) and the $\wt{{\rm Li}}_3$ solution with 
$z=2$ for large $r$ (UV) .  The plot of $\pr_{\rho}F$ shows 
that it starts from $1$ and goes to the value $2$, the other 
plots show how fluxes of the gauge fields grow, approaching 
constant values at large $r$. 

\subsubsection{Flows from AdS$_5$}

The flows which approach the asymptotically AdS$_5$ solution 
in the UV and end at AdS$_3$ or Li$_3$ in IR are easy to 
construct numerically, integrating outward from the IR. 
We find that the endpoint of the  flow from AdS$_5$ is 
uniquely determined by the pair $\left\{Q,\l\right\}$, 
where $\l$ is the coefficient in front of the slow fall-off 
mode in the expansion of the 5D dilaton field near the 
AdS$_5$ solution, 
\be
\varphi = \fr{1}{\sqrt{2}} + \frac{\l}{r^2}\ln{r} +\frac{\eta}{r^2} + \dots.
\label{dilaton-expansion}
\ee
On the field theory side, $\l$ corresponds to the source of 
an operator $\mathcal{O}_2$, as discussed in Maldacena and Nunez
\cite{Maldacena:2000mw}, however, for future reference we note
that the deformation parameter used there, $\bar\lambda$, 
is related to our $\lambda$ via
\be
\bar{\lambda} = \frac{\sqrt{2}}{3} e^{2h_0} \lambda
\label{lambdarel}
\ee
This operator (together with the curvature of the $\mathcal H_2$ 
and the flux $Q$) induces the RG flow on the field theory side. 
As noted previously, the fact that these flows only involve 
turning on a source for this operator implies that the flows 
to Lifshitz spacetimes break the Lorentz invariance spontaneously. 

\begin{figure}
\centering
\includegraphics[scale=0.7]{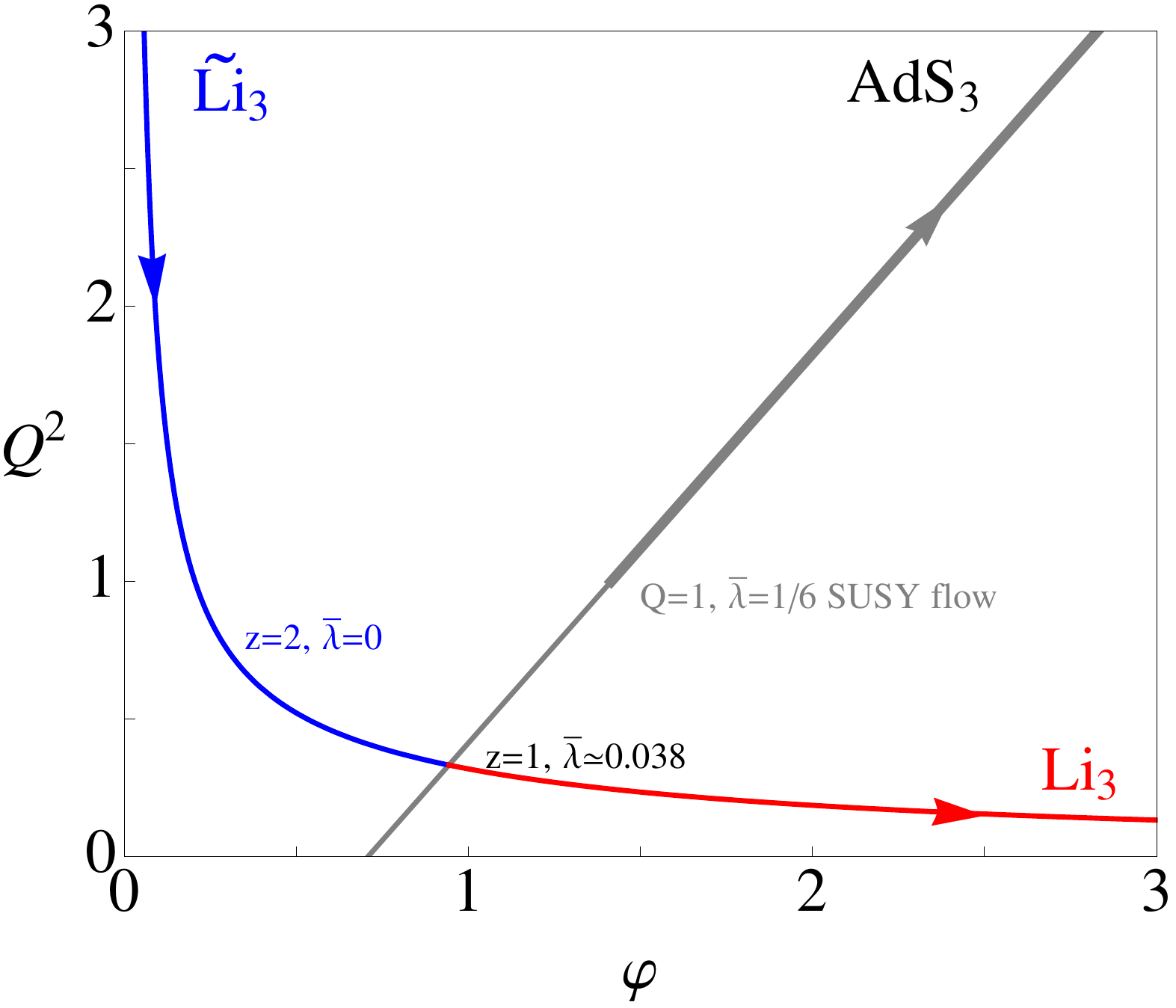}
\caption{Plots of AdS$_3$, $\wt{Li_3}$ and Li$_3$ solutions, 
indicating the corresponding value of  ${\bar\lambda}$ in the asymptotically 
AdS$_5$ UV region in the flow solutions. The arrows indicate
the direction of increasing ${\bar\lambda}$.}
\label{sol-plot-phi-AdS5}
\end{figure}

The values of ${\bar\lambda}$ for which we flow to the different 
solutions are presented schematically in Figure~\ref{sol-plot-phi-AdS5}. 
If we move along the AdS$_3$ (grey) line in the direction 
of increasing of $Q$, then the corresponding value of ${\bar\lambda}$ 
is also increasing. For $Q=0$ ${\bar\lambda}=0$, while for $Q=1$ 
${\bar\lambda}=\fr16$; this latter value corresponds to the supersymmetric 
flow of \cite{Maldacena:2000mw}. If we move along the 
$\wt{Li_3}$ (blue) line up (in the direction of increasing  
$Q$ and also increasing $z$), then the corresponding 
value of ${\bar\lambda}$ is decreasing, in such a way that for $Q=\sqrt{\fr23}$ 
($z=2$) ${\bar\lambda}=0$.\footnote{This is a numerical result, but it 
seems very reasonable, because in Lifshitz theories, 
a theory with $z=2$ always was a special case.}  
Above this point ${\bar\lambda} < 0$. If we move along the Li$_3$ (red) 
line down (in the direction of decreasing $Q$, but increasing  
$z$), then the corresponding value of ${\bar\lambda}$ is increasing. 
Numerically, ${\bar\lambda} \to \fr16$ as $z \rightarrow 2$ 
($Q \to 0$). 
We will discuss the field theoretic implications of the
values of ${\bar\lambda}$ in the next section, but first comment
on stability of the supergravity solutions.

\subsection{Stability to condensation of supergravity fields} 

In the analysis of the linearized perturbations, we encountered 
some complex eigenvalues for some values of parameters, 
as in the analysis of the IIA case in \cite{Braviner:2011kz}. 
For a decoupled scalar, such complex eigenvalues appear 
when the scalar violates the Breitenlohner-Freedman bound, 
and there is then an instability to condensation of the scalar. 
We would expect that there will be a similar instability to 
condensation of the modes with complex eigenvalues in 
our case, although we will not attempt to carry out a 
time-dependent analysis to demonstrate this instability explicitly. 
Certainly the appearance of the complex eigenvalues 
obstructs the usual interpretation of the eigenvalue as the 
dimension of the corresponding operator in the field theory. 

Also, it was noted in \cite{Amsel:2009ev} that purely from a 
bulk spacetime perspective, when such complex eigenvalues 
appear for a scalar field there is no boundary condition 
which preserves  the inner product which is invariant under 
the Lifshitz scaling isometry. Thus, we expect that in the cases 
with complex eigenvalues, we simply cannot choose boundary 
conditions such that our bulk solution is dual to an anisotropic 
scaling invariant field theory with a conserved inner product. 

A nice field theory dual description of the fixed points with 
complex eigenvalues is thus unlikely to exist. This leaves 
as potentially interesting cases a range of the AdS$_3$ fixed points 
and a range of the untilded  Li$_3$ fixed points with $z$ 
near 2. This is an interesting range of Lifshitz solutions, 
and an improvement of the IIA case, where the Lifshitz 
solutions with no complex eigenvalues were at larger values of $z$. 

\section{The UV field theory}
\label{stab}

Our interest in studying flows, particularly those from 
asymptotically AdS$_5$ spacetimes, is mainly that they might 
help us to understand the field theories dual to these spacetimes. 
In this section, we consider some stability issues that can 
obstruct our ability to learn about the field theory from these flows. 
For field theory on a flat space, the scalars in the adjoint 
of $SU(N)$ have flat directions corresponding to the Coulomb branch. 
However in our class of spacetimes, we are 
compactifying two of the directions on which the field theory lives on 
a space of negative curvature. One might therefore expect 
the curvature coupling of the field theory scalars to produce 
a runaway instability for the diagonal components of these 
scalar matrices. From the bulk spacetime point of view, the 
diagonal components of the scalars are positions of branes, 
so this runaway would be a brane nucleation instability. 

The story is of course more complicated, because in addition 
to the negative curvature space, we are introducing a flux 
$F^{(3)}_{y_1 y_2} = q/y_2^2$ on these directions, and also 
adding a source for the operator dual to the 5D dilaton $\phi$. 
In the supersymmetric case analysed in \cite{Maldacena:2000mw}, 
the effects of these deformations combine to preserve a 
twisted supersymmetry. The whole RG flow is supersymmetric, 
so on the field theory side the deformation of $\mathcal N = 4$ 
SYM is preserving some supersymmetry. One would then not 
expect the field theory to have a scalar instability, and indeed 
the terms combine to leave us with flat directions for some of 
the field theory scalars \cite{Maldacena:2000mw}. 
Similarly, from the bulk perspective, the addition of
the flux and deformation of the S$^5$ (encoded in the 5D dilaton)
will modify both the DBI and WZ components of a probe brane
action, which could stabilise the brane.

We now present analyses from both points of view -- using the
Maldacena-Nunez approach to contruct the field theory, then
confirming our results by a direct probe brane calculation.

\subsection{UV field theory analysis}

Let us analyze the field theory deformation for our general 
family of flows. The field theory includes six real scalars, 
transforming in the vector representation of the $SO(6)$ 
R-symmetry group and the adjoint of $SU(N)$. The 
consistent truncation we work with preserves an 
$SU(2) \times U(1)$ subgroup of $SO(6)$, so it is 
convenient to organize the scalars into three complex 
scalar fields $\W_1, \W_2$ and $\W_3$, where $\W_1$ 
and $\W_2$ transform under the $SU(2)$ and $\W_3$ 
transforms under the $U(1)$. The bulk 5D dilaton $\phi$ 
corresponds to an operator $\mathcal{O}_2$ which is 
a symmetric traceless combination of the scalars  transforming 
in the $\bf 20$ of $SO(6)$ \cite{Maldacena:2000mw}, 
\be
\mathcal{O}_2 = Tr \left\{ \fr23 \left|\W_3\right|^2 
- \fr13 \left( \left|\W_1\right|^2 + \left|\W_2\right|^2 \right) \right\}.
\label{operatorO2}
\ee
The deformation we consider has a negative curvature in 
the $y_1, y_2$ directions and a flux of the $\tau^3$ component 
of the $SU(2)$ gauge field through those directions, and a 
source for $\mathcal{O}_2$ with a coefficient $\bar\lambda$.
This corresponds to a deformation of the scalar part of 
the field theory Lagrangian to 
\be
S = \int d^4x \left\{ \fr12 \left|D_{\mu}\W_1\right|^2 
+ \fr12 \left|D_{\mu}\W_2\right|^2 + \fr12 \left|\pr_{\mu}\W_3\right|^2 
- \fr{R}{12} \sum_i \left|\W_i\right|^2 
+ \fr34 {\bar\l} R \mathcal{O}_2 \right\},
\label{scal-action}
\ee
where $D_{\mu} = \pr_{\mu} + \I A_{\mu}$ is the 
gauge-covariant derivative with respect to the component 
of the $SU(2)$ gauge field we turn on, and $R$ is the 
Ricci scalar of the two dimensional hyperbolic spacetime 
(note $R = - \left|R\right| < 0$). 
Substituting in $A_{y_1} = q/y_2$, we have 
\be
\beal
S = \int d^4x \Biggl\{ & \fr12 \sum_i \left|\pr_{\mu}\W_i\right|^2 
- \left|R\right| \left( \fr{\bar\lambda}{2} 
- \fr{1}{12} \right) \left|\W_3\right|^2 \\
&-\left|R\right| \left[ \frac{Q^2}{8} 
- \left(\fr{\bar \lambda}{4} + \fr{1}{12} \right) 
\right] \left( \left|\W_1\right|^2 + \left|\W_2\right|^2 \right) \Biggr\},
\eeal
\label{FTaction}
\ee
where the normalization of the $Q^2$ term and the coefficient 
of $\bar\l$ have been fixed by reference to the supersymmetric 
case, which corresponds to ${\bar\l} = \fr16$ and $Q=1$.

\subsection{Probe brane calculation}

We now want to explore this field theory from the bulk perspective.
Holographically, R-symmetry scalar fields correspond to 
inserting a brane with its four infinite dimensions parallel
to an $r=$const.\ section of the 5D space, and at a given
position on the (possibly distorted) $S^5$.
The effective action of such a probe brane is given
by the sum of a geometric DBI term, and a topological 
WZ term:
\be
S = - T_3 g_s^{-1} \int e^{-\Phi}
\sqrt{-{\rm det}[\gamma_{AB}+{\tt{F}}_{AB}]} d^4 \zeta 
+ T_3 \int C_4 
\ee
where $\zeta^A$ are the intrinsic coordinates on the brane
worldvolume; $\gamma_{AB}$ the induced metric;
${\tt{F}}_{AB}={\tt{B}}_{AB} + 2 \pi \alpha'F_{AB}$,
the pullback of the 2-form field to the
brane (zero in this background) and worldvolume 
gauge field (which we also set to zero); finally, $C_4$ is the
pullback of the 4-form gauge potential onto the brane.

In order to compute this action, we first need the background
geometry.
The twisting introduced previously corresponds to a distortion
of the $S^5$ in the reduction of the IIB SUGRA as 
described in \cite{Lu:1999bw}\footnote{Note that there are
some factors of two between the variables used here
and those of \cite{Lu:1999bw}: $(\phi)_{LPT}=\phi/2$, 
$(g_i)_{LPT}=g_i/2$, and $A_{LPT}=2A$, where 
$A$ stands for either the U(1) or SO(3) gauge field.}.
Lifting the 5D solutions of (\ref{metric},\ref{gauge})
to 10D, and writing
\be
\beal
S&=\sin\chi \;\;\;\;& \Delta &= \xi^2 S^2 + \xi^{-1} C^2 \\
C&=\cos\chi & U&=\xi S^2 + \xi^{-2} C^2 +\xi
\eeal
\label{SCDU}
\ee
gives\footnote{We have set $g_1=g_2/\sqrt{2}=2$ to
match the conventions of \cite{Maldacena:2000mw}} 
\cite{Gregory:2010gx}:
\be
\beal
ds^2 & = \Delta^{\fr12} \left( e^{2F} dt^2 - r^2 dx^2 
- e^{2d} \fr{dr^2}{r^2} - e^{2h} \fr{dy_1^2 + dy_2^2}{y_2^2} \right) \\
& - \xi^{-1} \Delta^{-\fr12} \left[ \Delta d\chi^2 
+ \xi^{-1} S^2 \left(d\eta - 2 \mathcal{A} \right) 
+ \fr14 \xi^2 C^2 \sum_i \left(h^{(i)}\right)^2 \right]
\eeal
\ee
\be
\beal
{\bf F_5} &=  2\, U \epsilon_5 + 3\, S\, C\, \xi^{-1} 
\star_5 d \xi \wedge d\chi
+ \frac{C^2}{2\sqrt{2}} \xi^2 \star_5 F^{(3)}_2
\wedge \sigma^{(1)} \wedge \sigma^{(2)} \\
& - \frac{S\,C\,}{\sqrt{2}}\xi^2 \star_5 F^{(3)}_2 \wedge h^{(3)}
\wedge d\chi -2 SC \xi^{-4}\star_5 {\cal F}_2 \wedge d\chi 
\wedge (d\eta - 2 {\cal A}) \,,
\eeal
\ee
the other form fields, the string dilaton and axion vanish.
Here, $h^{(i)}$ are the left invariant forms on $S^3$ 
($\sigma^{(i)}$) modified by the $SO(3)$ gauge fields:
\be
h^{(i)} = \sigma^{(i)} -  2\sqrt{2} \,A^{(i)} \, .
\ee
For constant $\xi$, we may reparametrize
the squashed $S^5$ as
\be
\beal
W_1 & = \xi \cos\chi \; \cos\frac{\theta}{2} 
\; e^{\I \fr{\phi+\psi}{2}} \\
W_2 & = \xi \cos\chi \; \sin\frac{\theta}{2} 
\; e^{\I \fr{\phi-\psi}{2}} \\
W_3 & = \xi^{-1/2} \sin\chi \;  e^{\I \eta}
\eeal
\label{positions}
\ee
which, together with the obvious definitions of the gauge
covariant differentiation for $W_{1,2}$ and $W_3$ give the
metric of the additional dimensions as
\be
ds_5 = 
- \xi^{-1} \Delta^{-\fr12} \left[\  |D W_1|^2 
+ |D W_2|^2 + |{\cal D} W_3|^2 \ \right]
\ee

As $\xi$ changes from unity, we can see how the $S^5$
becomes distorted while maintaining an $SO(3)\times U(1)$ symmetry.
Our 5D dilaton is thus a shape modulus for the $S^5$.
Since $\xi\equiv1$ for AdS$_5$, it is now transparent how to 
deal with the degrees of freedom of the probe brane: we simply
replace the `$\xi$' in \eqref{positions} with a radial variable
$r(\zeta)$, and allow the remaining angular degrees of freedom 
of the brane to also depend on the brane coordinates $\zeta^A$.
We will then expand the action for a slowly moving brane
at large $r$ in the asymptotic AdS$_5$ solution. 

We start with the  DBI part of the action
\be
S_{DBI} \propto - \int d^{4}\zeta \sqrt{-\det \gamma_{AB}}
\ee
where
\be
\gamma_{AB} = \fr{\pr X^a}{\pr\zeta^{A}} 
\fr{\pr X^b}{\pr\zeta^{B}} g_{ab}
\ee
with $X^{\mu} = \left[ t, x, r(\zeta), y_1, y_2, \chi(\zeta), 
\eta(\zeta), \theta(\zeta), \phi(\zeta), \psi(\zeta) \right]$ being
the brane's spacetime coordinates in terms of the intrinsic
coordinates $\zeta$, for which we choose the gauge
$\zeta^{A} = \left( t, x, y_1, y_2 \right)$. Thus
\be
\gamma_{AB} = \gamma^0_{AB}-\frac{1}{r^2} \left [
D_A W_1 \overline{D_B W_1}
+ D_A W_2 \overline{D_B W_2}
+ {\cal D}_A W_3 \overline{{\cal D}_B W_3}\;\right]
\ee
where $\gamma^0_{AB} = \Delta^{\fr12} \cdot 
\mbox{diag}\left(e^{2F},-r^2, -\fr{e^{2h}}{y_2^2}, 
-\fr{e^{2h}}{y_2^2} \right)$, the $1/r^2$ factor
arising because we have replaced $\xi$ with $r$ in 
\eqref{positions}. Hence, 
\beq
\sqrt{-\det\,\gamma_{AB}}\simeq
\sqrt{-\det\,\gamma^0_{ab}} \left( 1 - \frac{1}{2r^2} \gamma^{0 AB} 
D_A W_i \overline{D_B W_i} \right)
\eeq
(where we understand the covariant derivative in the sum
to be the one relevant to the particular $W_i$).
Since we are only interested in the leading order behaviour
as we change $W_i$, we only require $\gamma^{0 AB}$ to leading
order in $W_i$, i.e.\ at the AdS$_5$ limit:
\be
\gamma^{0 AB}\big |_{AdS5} = \fr{1}{r^2} \cdot 
\mbox{diag}\left( 1,-1, -y_2^2 e^{-2h_0}, -y_2^2 e^{-2h_0} \right)
\ee
hence 
\beq
S_{DBI} \propto -\int d^{4}\zeta \; \frac{r \Delta}{y_2^2} e^{F+2h} 
\left( 1 - \fr{1}{2r^4} \sum_i \left|D_{\mu} W_i\right|^2 \right)
\eeq

For the WZ term, note that although the 4-form potential is
rather involved for a general flow, we only require the leading
order part parallel to the probe brane worldvolume, which can
be found by integrating the $U$ function in \eqref{SCDU}.
Putting this together, we see that
\beq
S_{\rm eff} \sim \int d^4\zeta \Biggl\{ -\Delta(\xi,\chi) 
\cdot r e^{F+2h} \left( 1 - \fr{1}{2r^4} \sum_i 
\left|D_{\mu} W_i\right|^2 \right) + 
2 \int e^{F+d+2h} U(\xi,\chi) dr \Biggr\}
\label{eff_action}
\eeq

We now expand this action in the asymptotic AdS$_5$ region, 
but with one difference to the procedure followed in 
\S \ref{ads5exp}: we need to consider a linear expansion 
in the case of {\em{finite}} volume of the 2D hyperbolic 
space, i.e.\ finite $h_0$. The full asymptotic solution 
together with corrected expansion up to $r^{-2}$ order reads
\be
\beal
F & = \ln{r}\;, \quad &
d & = - \fr{e^{-2h_0}}{6r^2}\;, \\
h & = \ln{r} + h_0 + \fr{e^{-2h_0}}{4r^2}\;, \quad &
\xi & = 1 + \fr{\sqrt{2}}{3}\, \frac{\lambda\ln{r}}{r^2} 
+ \fr{\sqrt{2}}{3} \fr{\mu}{r^2}\;. 
\eeal
\ee

Substituting these expressions into (\ref{eff_action}), and 
performing the integral for $U$, we see that 
all terms proportional to $\mu$ and $\lambda\ln r$ cancel
leaving 
\be
S_{\rm eff} \sim \int d^4 \zeta \left\{ \fr12 e^{2h_0} \sum_i 
\left|D_{\mu} W_i\right|^2 - \fr{\lambda}{3\sqrt{2}} e^{2h_0} 
\left(2S^2 - C^2\right) r^2 + \fr{1}{6} r^2 \right\}
\label{seff1}
\ee
It is easy to see that we can identify
\be
\left(2S^2 - C^2\right) r^2 = 3 \mathcal{O}_2\;, 
\quad r^2 = \sum_i \left|W_i\right|^2
\ee
and noting the relation between our $\lambda$ and $\bar\lambda$,
\eqref{lambdarel}, as well as the curvature of the 2D hyperbolic 
space,  $R = - 2 e^{-2h_0}$, we get
\be
S_{\rm eff} \propto \int d^4 \zeta \; e^{2h_0} 
\left\{ \fr12 \sum_i \left|D_{\mu}W_i\right|^2 
- \fr34 {\bar\lambda} R \mathcal{O}_2 + \fr{1}{12} 
R \sum_i \left|W_i\right|^2  \right\}
\ee
which coincides with the expression for the field 
theory effective action (\ref{scal-action}) precisely
\footnote{Indeed, the uplift of the AdS flows can be
generalised in the context of solutions in $D=10,11$
dual to ${\cal N}=2$ SCFT's, as studied in
\cite{Gauntlett:2007ma,Gauntlett:2007sm}.
(We thank Jerome Gauntlett for pointing this out.)}.

\subsection{Stability and Lifshitz dual field theories}

Having obtained the field theory action, \eqref{FTaction},
we now analyse the scalar stability. In order to have 
stable potential for the $\W_3$ field, we should have
\be
\fr12 {\bar\l} - \fr{1}{12} \geq 0 \Rightarrow {\bar\l} \geq \fr16,
\ee
While for the twisted fields $\W_1$ and $\W_2$ we should have
\be
\fr{Q^2}{8} - \left(\fr14 {\bar\l} + \fr{1}{12} \right) \geq 0.
\label{stability-twisted}
\ee
For the supersymmetric case, both these bounds are 
automatically saturated (by our choice of normalization 
in matching operator sources to bulk modes), reproducing 
the flat directions of \cite{Maldacena:2000mw}. 

For AdS$_3$ solutions we know that in the AdS$_3$ region 
$Q^2 = \varphi \sqrt{2} - 1$, and, by numerical analysis we 
determine ${\bar\lambda}$ as a function of the value of $\varphi$ 
in the AdS$_3$ region. The stability criterion for the $\W_3$ 
field, ${\bar\lambda} \geq 1/6$, which corresponds to 
$\varphi \geq \sqrt{2}$. Meanwhile, \eqref{stability-twisted}
provides an upper bound on $\varphi$, as $\bar\l$ increases more
rapidly than $Q^2$ along the family of AdS$_3$ flows. Numerically,
we find that the AdS$_3$ solutions with 
\be
\varphi \in \left[ \sqrt{2} , \sim 3.26 \right]
\ee
result from an RG flow from a field theory in the UV where the 
field theory deformation is not introducing a field theory scalar 
instability. The corresponding region for the charge $Q$ is
\be
Q^2 \in \left[ 1 , \sim 3.61 \right]\,.
\ee

Disappointingly, for the Lifshitz solutions we found numerically 
that none of the solutions involve flows with ${\bar\lambda} \geq 1/6$. 
The flows on the untilded branch do approach ${\bar\lambda} \to 1/6$ 
when $z \to 2$, but  $Q \to 0$ in this limit, 
so even if we are nearly satisfying the stability condition for 
$\W_3$ in the limit, the condition for $\W_1$ and $\W_2$ is 
badly violated. Thus, none of our Lifshitz solutions is obtained 
as an RG flow from a stable UV field theory, and we cannot 
use these  RG flows to define the field theory dual to the IR fixed points.

This UV instability does not necessarily imply that the IR fixed 
points are ill-defined, just that this approach to constructing 
them has failed. There are solutions on the Li$_3$ branch 
for which we did not have evidence of a supergravity instability 
which are still candidates for having a dual field theory; but we 
will have to look elsewhere for a top-down definition of this field theory.

\section*{Acknowledgements} We are grateful for useful 
conversations with Mukund Rangamani, and for collaboration 
with Ludovic Plante on an early version of this work. PB would 
like to thank Perimeter Institute for hospitality.
PB is supported by an EPSRC International Doctoral Scholarship,
RG and SFR are supported in part by STFC (Consolidated Grant ST/J000426/1).
RG is also supported by the Wolfson Foundation and Royal Society, and
Perimeter Institute for Theoretical Physics. 
Research at Perimeter Institute is supported by the Government of
Canada through Industry Canada and by the Province of Ontario through the
Ministry of Research and Innovation.  

\appendix
\section{Appendix A: Additional AdS Solutions}

In the main text we assumed that the topologically charged 
part of the fluxes, i.e.\ the flux through the compact hyperbolic 
space, only involved the $SU(2)$ gauge field, as this is the only 
possibility for the Lifshitz solutions \cite{Gregory:2010gx}. However, more 
generally the abelian field could also have a topological flux. 
Here we will briefly discuss constructing more general 
AdS$_3$ geometries using this freedom. These solutions were also obtained in a more systematic analysis in \cite{Benini:2012cz,Benini:2013cda}.

Introducing the following more general ansatz for the gauge fields
\bea
\mathcal{F}_{y_1y_2}  = \fr{q_1}{y_2^2}, \\
F^{^(3)}_{y_1y_2}  = \fr{q_2}{y_2^2}, \nn
\eea
together with the standard ansatz for the metric (\ref{metric}) 
with $r$-independent constants $d_0$ and $h_0$ and 
$F(\rho) = \rho$, gives rise to the following system of equations
\bea \label{additional-ads}
2 e^{-2D_0} & = & \fr16 \left(\varphi_0^{-\fr23} 
+ 2\sqrt{2}\varphi_0^{\fr13}\right) + \fr23 \varphi^{\fr23} Q_2^2 e^{-4H_0} 
+ \fr23 \varphi^{-\fr43} Q_1^2 e^{-4H_0}, \\
e^{-2H_0} & = & \fr16 \left(\varphi_0^{-\fr23} 
+ 2\sqrt{2}\varphi_0^{\fr13}\right) - \fr43 \varphi_0^{\fr23} Q_2^2 e^{-4H_0} 
- \fr43 \varphi_0^{-\fr43} Q_1^2 e^{-4H_0}, \nn\\
0 & = & \fr12 \left(-\varphi_0^{-\fr23} + \sqrt{2}\varphi_0^{\fr13}\right) 
- 2 \varphi_0^{\fr23} Q_2^2 e^{-4H_0} 
+ 4 \varphi_0^{-\fr43} Q_1^2 e^{-4H_0}, \nn
\eea
where $Q_1 = q_1 g_1$. Solving this system gives us a 
two-parameter family of AdS$_3$ solutions,
\bea
e^{-2D_0} & = & f_D \left(Q_1,Q_2\right), \\
e^{-2H_0} & = & f_H \left(Q_1,Q_2\right), \nn\\
\varphi_0 & = & f_{\varphi} \left(Q_1,Q_2\right), \nn
\eea
which will coincide with (\ref{AdS3 solution}) if we put $Q_1 = 0$, 
$g_1 = 2$, $g_2 = 2\sqrt{2}$ and $Q_2 = Q$. 
These solutions are supersymmetric if 
\be
Q_1 + Q_2 = 1.
\ee
Field theory duals for two points in this family ($Q_1=1$ and $Q_2=1$) 
were discussed through twisting in \cite{Maldacena:2000mw}. 
There it was also pointed out that the field theory description 
of the general supersymmetric solution of (\ref{additional-ads}) 
would involve some fields acquiring fractional spins during twisting.

\vskip 5mm
\break

\providecommand{\href}[2]{#2}
\begingroup\raggedright\endgroup
\end{document}